\documentclass[12pt]{article}
\usepackage{amsmath,amssymb}
\usepackage{latexsym,graphicx,color}

\def\6#1{{\underline{#1}}}
\def\m6#1{{\underline{#1}\,}}

\newdimen\Tdim
\def\ispan{{\setbox0=\hbox{i}%
\Tdim\ht0\advance\Tdim\dp0\rule[-\dp0]{0pt}{\Tdim}}}
\def\jspan{{\setbox0=\hbox{j}%
\Tdim\ht0\advance\Tdim\dp0\rule[-\dp0]{0pt}{\Tdim}}}
\def\Tspan#1{{\setbox0=\hbox{#1}%
\Tdim\ht0\advance\Tdim\dp0\advance\Tdim.55ex\rule[-\dp0]{0pt}{\Tdim}\box0}}

\def\be{\begin{eqnarray}}
\def\ben{\begin{eqnarray*}}
\def\ee{\end{eqnarray}}
\def\een{\end{eqnarray*}}
\def\Tr{{\rm Tr}}

\def\p{\partial}

\def\D{\mathcal{D}}

\def\=:{=\hspace{-.7em}\raisebox{1.1ex}{.}\hspace{.1em}\raisebox{-0.2ex}{.} }

\newcommand {\beq}{\begin{eqnarray}}
\newcommand {\eeq}{\end{eqnarray}}

\makeatletter
\@addtoreset{equation}{section}

\renewcommand{\thefootnote}{\fnsymbol{footnote}}
\makeatother

\makeatletter
\newcommand{\thetablename}{Table}
\def\fnum@table{\thetablename\ \thetable}
\makeatother

\begin{document}
\thispagestyle{empty}
\begin{flushright}
IFUP-TH/2007-29\\
November, 2007 \\
\end{flushright}
\vspace{3mm}

\begin{center}
{\LARGE \bf
Static Interactions of non-Abelian Vortices} \\
\vspace{5mm}

{\normalsize\bfseries Roberto~Auzzi$^{a}$, \ Minoru~Eto~$^{b,c}$ \ and\ Walter~Vinci~$^{b,c}$ \footnotetext{ e-mail:
\tt r.auzzi(at)swan.ac.uk, minoru(at)df.unipi.it, walter.vinci(at)pi.infn.it }

\vskip 1.5em
$^a$ {\it Department of Physics, Swansea University, Singleton Park, Swansea SA2 8PP, U.K.
}
\\
$^b$ {\it INFN, Sezione di Pisa,
Largo Pontecorvo, 3, Ed. C, 56127 Pisa, Italy
}
\\
$^c$ {\it Department of Physics, University of Pisa,
Largo Pontecorvo, 3,   Ed. C,  56127 Pisa, Italy
}}
\vspace{12mm}

\abstract{ Interactions between non-BPS non-Abelian vortices are
studied in non-Abelian $U(1) \times SU(N)$ extensions of the Abelian-Higgs model in four dimensions.
The distinctive feature of a non-Abelian
vortex is the presence of an internal ${\bf C}P^{N-1}$ space
of orientational degrees of freedom.
For fine-tuned values of the couplings, the vortices are BPS
and there is no net force between two static parallel vortices at arbitrary distance.
On the other hand, for generic values of the couplings
the interactions between two vortices
depend non-trivially on their relative internal orientations.
We discuss the problem both with a numerical approach (valid for small deviations from the BPS limit) and in
a semi-analytical way (valid at large vortex separations).
The interactions can be classified with respect
to their asymptotic property
at large vortex separation. In a simpler fine-tuned model, we find two regimes which are quite similar to the usual type
I/II Abelian superconductors. In the generic model we find other two new regimes: type I$^*$/II$^*$. Unlike the type I
(type II) case, where the interaction is always attractive (repulsive), the type I$^*$ and II$^*$ have both attractive
and repulsive interactions depending on the relative orientation. We have found a rich variety of interactions at small
vortex separations. For some values of the couplings, a bound state of two static vortices at a non-zero distance
exists. }

\end{center}

\vfill
\newpage
\setcounter{page}{1}
\setcounter{footnote}{0}
\renewcommand{\thefootnote}{\arabic{footnote}}

\section{Introduction}

Nobody doubts the importance of topological solitons in various areas of modern physics (see \cite{Manton:2004tk} for a
general review). They are closely related to the phenomena of spontaneous symmetry breaking which e.g. occur as a phase
transition from the high temperature phase of the early universe to the present cold universe.
In particular, vortex strings are believed to play important roles in the confinement of quarks in QCD, and
they could be relevant to the study of cosmic string effects in the early universe. Historically, the vortex string, as
a topological soliton, was found in the Abelian-Higgs model by Abrikosov and Nielsen-Olesen
\cite{Abrikosov:1956sx,Nielsen}.

Recently, a new type of vortex was found in $U(N)$ non-Abelian gauge theories coupled with $N_f=N$ Higgs fields in the
fundamental representation \cite{Hanany:2003hp,Auzzi:2003fs}. This object is called non-Abelian vortex. A typical
feature of this vortex is that it possesses internal degrees of freedom, which arise when the vortex breaks an exact
flavor symmetry of the vacuum.
These degrees of freedom are related to the orientations of the non-Abelian flux inside the vortex core.
The vacuum of the theory leaves the colour-flavour locked $SU(N)_{{\rm C}+{\rm F}}$ symmetry;
on the other hand, the vortex
soliton spontaneously breaks this $SU(N)_{{\rm C}+{\rm F}}$ symmetry to
$SU(N-1)_{{\rm C}+{\rm F}}\times U(1)_{{\rm C}+{\rm F}}$;
these broken symmetries give rise to the moduli space of an elementary vortex
\[ {\bf C}P^{N-1}=\frac{SU(N)_{{\rm C}+{\rm F}}}{SU(N-1)_{{\rm C}+{\rm F}}\times U(1)_{{\rm C}+{\rm F}}}.\]
The classical moduli coordinate can be promoted to a field living in the vortex worldvolume; in this way
vortex solitons in a $3+1$ dimensional theory are directly connected with a ${\bf C}P^{N-1}$ sigma
model in $1+1$ dimension, which describes the macroscopic physics of the flux tube. Several groups have
intensively investigated these objects in relations to various aspects of physics; a partial list includes confined
monopoles \cite{monop}, quantum aspects \cite{nsgsy,ssize,tongheterotic}, higher winding numbers
\cite{Eto:2005yh,Auzzi:2005gr,Eto:2006cx}, relation to D-branes in string theory \cite{Hanany:2003hp,abe},
dualities \cite{seibvortex,Gorsky:2007ip,Eto:2006dx},
cosmic strings \cite{Hashimoto:2005hi,Eto:2006db}, semilocal extensions
\cite{Vachaspati:1991dz,Shifman:2006kd}, $SO(N)$ generalization \cite{Ferretti:2007rp}, high temperature QCD
\cite{Gorsky:2007kv}, global vortices \cite{global}, gravity \cite{Aldrovandi:2007bn},
composite states of various BPS solitons \cite{monop,composite}
and statistical mechanics \cite{Eto:2007aw}.
Readers can find good
reviews in \cite{reviews,Eto:2006pg}.

Most the works on the non-Abelian vortex so far
 were focused on the BPS limit~\cite{Bogomolny:1975de}
(a single non-Abelian non-BPS vortex configuration is discussed
in \cite{mmy,Bolognesi:2004da,seibvortex,Gorsky:2007kv}).
No forces arise among BPS vortices, because there is a nice
balance between the repulsive forces mediated by the vector particles and attractive forces mediated by the scalar
particles. In this particular limit, the solutions
to the equations of motion develop a full moduli space of solutions \cite{Manton:1981mp}.
However, once the balance between the attractive force and the repulsive force is lost,
 the moduli space disappears. Alternatively
we can think that an effective potential is generated on this moduli space. It is well known that ANO vortices in the
type I system feel an attractive force while those in the type II model feel a repulsive force
\cite{Bogomolny:1975de,Jaffe:1980mj,Gustafson:2000, Jacobs:1978ch,Bettencourt:1994kf,Speight:1996px}. In condensed
matter physics, it is also known that type II vortices form the so-called Abrikosov lattice,
\cite{Abrikosov:1956sx,Kleiner1964} due to the repulsive force between them. Furthermore, lattice simulations
give some evidence of the presence of a (marginal) type II superconductivity in QCD \cite{lattice}.

We are interested in studying interactions between non-Abelian vortices which are non-BPS.
In non-supersymmetric theories, BPS configurations
are obtained with fine-tuned values of the couplings.
If supersymmetry exists in the real world, it is surely broken at a low energy scale;
therefore non-BPS vortices are more natural than BPS ones.
Also, we encounter such non-BPS configurations in supersymmetric theories
\cite{Auzzi:2003fs,Eto:2006cx,Eto:2006dx}
when we consider a hierarchical symmetry breaking closely related to a dual
picture of color confinement of truly non-Abelian kind.
Specifically we are interested in the interactions between
vortices with different internal orientations, which is the distinct feature from the ANO case. In a previous
paper~\cite{1stpaper}, we have discussed these aspects in an $\mathcal{N}=2$ theory with an adjoint mass term which
breaks the extended supersymmetry, and we have found a natural non-Abelian generalization of type I superconductors.
Even if the force between two non-Abelian vortices is not always attractive, we have found a close resemblance with
type I Abelian vortices: the lightest field of the theory is a scalar field. So if we put two vortices at large
distance, the prevailing part of the interaction is mediated by the scalar particles and not by vector particles.
Moreover, if the two vortices have the same orientation in the internal moduli space, the force is always attractive.

In this paper we study the same problem in another theoretical setting: an extension of the Abelian-Higgs model
with arbitrary scalar couplings which is generically incompatible with the BPS limit. The simplest extension in this
direction is a theory in which there are just two mass scales; the mass of the vector bosons and the mass of the
scalars. There is one parameter $\lambda$ which controls the ratio of the two mass scales. We find that
$\lambda < 1$ leads to an attractive force as a usual Abelian type I, while for $\lambda > 1$ a repulsive force works,
similarly to the usual Abelian type II. There is no force between vortices with opposite ${\bf C}P^{N-1}$ orientation.
For $\lambda<1$ (type I) this configuration is unstable and the true minimum of the potential corresponds to two
coincident vortices with the same orientation. For $\lambda>1$ (type II) this configuration is stable; in other words a
part of the moduli space corresponding
to the relative distance between vortices with opposite orientations
survives the non-BPS perturbation.

However, in more general theories where the masses of the Abelian and non-Abelian degrees of freedom are different, we
find a more complicated picture. There are four mass scales, the masses of the $U(1)$ and of the $SU(N)$ vector bosons,
the masses of the scalars in the adjoint representation and the singlet of $SU(N)_{{\rm C}+{\rm F}}$. At large
distance, the interaction between two vortices is dominated by the particle with the lightest mass. So if we keep the
four masses as generic parameters,
 at large vortex separation we find four different regimes that we call Type I, Type II,
Type I$^*$ and Type II$^*$. In the last two categories
repulsive and attractive interactions depend on the relative orientation.
We study also numerically the interactions
among two vortices at any separation with arbitrary orientations, and find that short distance forces also have
rich qualitative features depending both on the relative orientations and the relative distance.

 The paper is organized as follows.
In Sect.~2 we describe the theoretical set-up. In Sect.~3 we write the equations for the vortex and we quickly review
the moduli space of the two vortices in the BPS limit. In Sect.~4 vortices in a fine-tuned setup are studied; the effective
vortex potential in the case of  small deviations from the BPS limit is found numerically. In Sect.~5 a more general
set-up with four independent parameters is discussed in the same way. In Sect.~6 the effective potential at large
vortex separation is found using a semi-analytical approach. Sect.~7 contains the conclusions. In the Appendix we
provide the link between the formalism of this paper and that of the companion paper~\cite{1stpaper}.

\section{Theoretical set-up}

\subsection{A fine-tuned model}
Our natural starting point is the following non-Abelian, $U(N)$, extension of the Abelian-Higgs model in four
dimensions:
\beq
{\cal L} =
\Tr\left[
- \frac{1}{2g^2} F_{\mu\nu}F^{\mu\nu}
+ \D_\mu H (\D^\mu H)^\dagger
- { \frac{\lambda^2 \, g^2}{4}} \left(v^2 {\bf 1}_{N} - HH^\dagger \right)^2
\right].
\label{eq:Lag_NAH}
\eeq
Here, for simplicity we take the same gauge coupling $g$ for both the $U(1)$ and $SU(N)$ groups, while $\lambda^2 \,
g^2/4$ is a scalar coupling and $v$ ($>0$) determines the Higgs VEV. In this simple model we have only three couplings
$(g,\lambda,v)$. The $N$ by $N$ matrix field $H$ embodies $N$ Higgs fields in the fundamental representation of $U(N)$.
There is also an $SU(N)$ flavor symmetry which acts on $H$ from the right hand side. The vacuum of the model is given
by:
\beq
HH^\dagger = v^2 {\bf 1}_N. \eeq
 The vacuum breaks completely the $U(N)$ gauge symmetry, although a global color-flavor locking symmetry
$SU(N)_{\rm C+F}$ is preserved
\beq
H \to U_{\rm G} H U_{\rm F},\quad U_{\rm G} = U_{\rm F}^\dagger,\quad
U_{\rm G} \in SU(N)_{\rm G},\ U_{\rm F} \in SU(N)_{\rm F}.
\eeq
The trace part $\Tr H$ is a singlet under the color-flavor group and the traceless parts are
in the adjoint representation.
We have two mass scales, one for the vector bosons and the other for the scalar bosons.
The $U(1)$ and the $SU(N)$ gauge vector bosons
have both the same mass
\beq
M_{U(1)}=M_{SU(N)}=g \, v .
\eeq
The masses of the scalars are given by the eigenvalues of the mass matrix. We start with $2 N^2$ real scalar fields in
$H$: $N^2$ of them are eaten by the gauge bosons (the Higgs mechanism) and the other $N^2$ (one singlet and
the rest adjoint) have same masses
\beq
M_{\rm s} = M_{\rm ad} = \lambda \, g \, v.
\eeq
When we choose the critical coupling $\lambda=1$ (BPS),
the mass of these scalars is the same as the mass of the gauge bosons and
the Lagrangian allows an $\mathcal{N}=2$ supersymmetric extension.
The BPS vortices saturating the BPS energy bound admit infinitely degenerate set of solutions.

\subsection{Models with general couplings}

A straightforward generalization of the fine-tuned model (\ref{eq:Lag_NAH}) is to consider
different gauge couplings, $e$ for the $U(1)$ part and $g$ for the $SU(N)$ part,
and a slightly more general scalar potential
\beq
{\cal L}
= \Tr \left[ - \frac{1}{2g^2} \hat F_{\mu\nu} \hat F^{\mu\nu} - \frac{1}{2e^2} f_{\mu\nu} f^{\mu\nu} + \D_\mu H (\D^\mu
H)^\dagger \right] - V, \label{flum}
\eeq
where we have defined $\hat F_{\mu\nu} = \sum_{A=1}^{N^2-1} F_{\mu\nu}^A T_A$ and $f_{\mu\nu} = F_{\mu\nu}^0 T^0$ with
$\Tr(T^AT^B) = \delta^{AB}/2$ and $T^0 = {\bf 1}/\sqrt{2N}$
\footnote{ In the case $g=e$ it is more compact to use
 $F_{\mu\nu} = \hat F _{\mu\nu} + f_{\mu\nu}$. }. The scalar potential is:
\beq
V &=& \frac{\lambda_g^2g^2}{2}\sum_{A=1}^{N^2-1} \left(H^{i\dagger}T^AH_i\right)^2 +
\frac{\lambda_e^2e^2}{4N}\left(H^{i\dagger}H_i - Nv^2\right)^2 \nonumber\\
&=& \frac{\lambda_g^2g^2}{4} \Tr \hat X^2 + \frac{\lambda_e^2e^2}{4}\Tr \left(X^0T^0 - v^2{\bf 1}_N\right)^2,
\label{eq:flum_pot}
\eeq
where
\beq
X \equiv HH^\dagger = X^0T^0 + \sum_{A=1}^{N^2-1}X^AT^A,\quad \hat X \equiv \sum_{A=1}^{N^2-1}X^AT^A = 2
\sum_{A=1}^{N^2-1} \left(H^{i\dagger}T^AH_i\right)T^A.
\eeq
The Lagrangian has the same symmetries as the previous fine-tuned model~(\ref{eq:Lag_NAH}). The potential in
Eq.~(\ref{eq:flum_pot}) is the most general gauge invariant quartic potential which can be built with the matter
content of the theory. The $U(1)$ and the $SU(N)$ vector bosons have different masses
\beq
M_{U(1)}=e \,v, \qquad M_{SU(N)}=g \, v.
\eeq
Moreover, the singlet part of $H$ has a mass $M_{\rm s}$ different from that of the adjoint part
$M_{\rm ad}$
\beq
M_{\rm s}=\lambda_e \, e \, v, \qquad M_{\rm ad}=\lambda_g \, g \, v.
\eeq
When we take equal couplings, $g=e$ and $\lambda\equiv\lambda_e=\lambda_g$, the scalar potential reduces to the
simple potential $V_{g=e} = \frac{\lambda^2g^2}{4} \Tr \left( X - v^2 {\bf 1}_N\right)^2$.
For the critical values $\lambda_e=\lambda_g=1$,
the Lagrangian allows an $\mathcal{N}=2$ supersymmetric extension and
then the model admits BPS vortices which saturate the BPS energy bound.


\section{Non-Abelian vortices in the fine-tuned model}
\subsection{Vortex equations}
We study the fine-tuned model (\ref{eq:Lag_NAH}) through out this section. For convenience, let us make the
following rescaling of fields and coordinates:
\beq
H \rightarrow v H,\quad
W_\mu \rightarrow gv W_\mu,\quad
x_\mu \rightarrow \frac{x_\mu}{gv}.
\eeq
The Lagrangian then in Eq.~(\ref{eq:Lag_NAH}) takes the form
\beq
\tilde {\cal L}=\frac{{\cal L}}{g^2 v^4} = \Tr\left[ - \frac{1}{2} F_{\mu\nu}F^{\mu\nu} + \D_\mu H (\D^\mu H)^\dagger -
\frac{\lambda^2}{4} \left({\bf 1}_{N} - HH^\dagger \right)^2 \right], \label{eq:reL}
\eeq
and the masses of vector and scalar bosons are
rescaled to
\beq
M_{U(1)} = M_{SU(N)} = 1,\qquad
M_{\rm s} = M_{\rm ad} = \lambda.
\eeq
As explained in the introduction,
the model with $\lambda <1$ ($\lambda>1$) in the Abelian case ($N=1$) is called type I (type II) and
the forces between vortices are attractive (repulsive).
At the critical coupling $\lambda =1$,  there are no forces
between vortices, so that multiple vortices stably coexist.

In order to construct non-BPS non-Abelian vortex solutions,
we have to solve the following 2nd order differential equations,
derived from the Lagrangian (\ref{eq:Lag_NAH}),
\beq
\D_\mu F^{\mu\nu} - \frac{i}{2}\left[H (\D^\nu H)^\dagger - (\D^\nu H) H^\dagger\right] = 0,
\label{eq:eom1}\\
\D_\mu\D^\mu H + \frac{\lambda^2}{4}\left(1-HH^\dagger\right)H = 0.
\label{eq:eom2}
\eeq
From now on, we restrict ourselves to static configurations
depending only on the coordinates $x^1,x^2$. Here we introduce a
complex notation
\beq
z = x^1 + i x^2,\quad
\p = \frac{\p_1-i\p_2}{2},\quad
W = \frac{W_1 - iW_2}2,\quad
\D = \frac{\D_1 - i\D_2}2 = \p + iW.
\eeq
The equation of motions are of course not gauge invariant but covariant. It might be better to study gauge invariant
quantities instead of dealing with the original fields $H$ and $W_\mu$. For this purpose we rewrite our fields as follows
\beq
\bar W (z,\bar z) = - i S^{-1}(z,\bar z)\bar\p S(z,\bar z),\quad
H(z,\bar z) = S^{-1}(z,\bar z) \tilde H(z,\bar z),
\label{eq:decomposition}
\eeq
where $S$ takes values in $GL(N,{\bf C})$ and it is in the fundamental representation of $U(N)$ while the gauge singlet
$\tilde H$ is an $N \times N$ complex matrix. There is an equivalence relation $(S,\tilde H) \sim (V(z)S,V(z)\tilde
H)$, where $V(z)$ is a holomorphic $GL(N,{\bf C})$ matrix with respect to $z$, because different elements in the same
equivalence class give us the same physical fields as in Eq.~(\ref{eq:decomposition}).
The gauge group $U(N)$ and the flavor symmetry act as follows
\beq
S(z,\bar z) \to U_{\rm G} S(z,\bar z),\quad
H_0(z) \to H_0(z) U_{\rm F},\qquad
U_{\rm G} \in U(N)_{\rm G},\
U_{\rm F} \in SU(N)_{\rm F}.
\eeq

In order to write down the equations of motion (\ref{eq:eom1}) and (\ref{eq:eom2}) in a gauge invariant fashion, we
introduce a gauge invariant quantity
\beq
\Omega (z,\bar z) \equiv S(z,\bar z) S(z,\bar z)^\dagger.
\eeq
With respect to the gauge invariant objects $\Omega$ and $\tilde H$, the equations (\ref{eq:eom1}) and (\ref{eq:eom2})
are written in the following form
\beq
4 \bar \p^2 \left( \Omega \p \Omega^{-1} \right) - \tilde H \bar\p \left( \tilde H^\dagger \Omega^{-1} \right)
+ \bar\p \tilde H \tilde H^\dagger \Omega^{-1} = 0 ,\qquad\quad
\label{eq:eom_omega1}\\
\Omega \p \left( \Omega^{-1} \bar \p \tilde H \right)
+ \bar \p \left( \Omega \p \left( \Omega^{-1} \tilde H \right)\right)
+ \frac{\lambda^2}{4} \left( \Omega - \tilde H \tilde H^\dagger \right) \Omega^{-1} \tilde H = 0.
\label{eq:eom_omega2}
\eeq
Notice that Eq.~(\ref{eq:eom_omega1}) is a 3rd order differential equation.
This is the price we have to pay in order to write
down the equations of motion in terms of gauge invariant quantities.
These equations must be solved with the following
boundary conditions for $k$ vortices:
\beq
\det \tilde H \rightarrow z^k,\quad \Omega \rightarrow \tilde H \tilde H^\dagger,
\qquad \text{as}\quad z \rightarrow \infty.
\eeq
 The field strength is given by
\beq
F_{12} = 2 S^{-1} \bar \p \left( \Omega \p \Omega^{-1} \right) S.
\eeq
Notice that Eq.~(\ref{eq:eom_omega1}) is invariant under the $SU(N)$ flavor symmetry while
Eq.~(\ref{eq:eom_omega2}) is covariant. This leads to Nambu-Goldstone zero modes for
vortex solutions.


\subsection{BPS Limit}

To see the relation with the BPS equations, let us take a holomorphic function $\tilde H$ with respect to $z$ as
\beq
\tilde H = H_0(z).
\eeq
Then the equations (\ref{eq:eom_omega1}) and (\ref{eq:eom_omega2}) reduce to
\beq
\bar \p \left[ 4 \bar \p \left(\Omega \p \Omega^{-1} \right) - H_0H_0^\dagger \Omega^{-1} \right]=0,\\
\bar \p \left( \Omega \p \Omega^{-1} \right)
+ \frac{\lambda^2}{4} \left( {\bf 1} - H_0 H_0^\dagger \Omega^{-1} \right) = 0.
\label{eq:master}
\eeq
These two equations are consistent only in the BPS limit $\lambda = 1$. The equation (\ref{eq:master}) is the master
equation for the BPS non-Abelian vortex and the holomorphic matrix $H_0(z)$ is called the moduli matrix
\cite{Eto:2005yh,Eto:2006pg}. For any given moduli matrix $H_0(z)$, given the corresponding solution to the master
equation, the physical fields $W_\mu$ and $H$ are obtained via Eq.~(\ref{eq:decomposition}).  All the complex
parameters contained in the moduli matrix are moduli of the BPS vortices. For example, the position of the vortices can
be read from the moduli matrix as zeros of its determinant $\det H_0(z_i) = 0$. Furthermore, the number of vortices
(the units of magnetic flux of the configuration) corresponds to the degree of $\det H_0(z)$ as a polynomial with
respect to $z$. The classification of the moduli matrix for the BPS vortices is given in
Ref.~\cite{Eto:2005yh,Eto:2006pg}.

From now on, we consider a $U(2)$ gauge theory ($N=2$), which is the minimal model for non-Abelian vortices. The
minimal winding BPS vortex is described by two moduli matrices
\beq
H_0^{(1,0)} =
\left(
\begin{array}{cc}
z-z_0 & 0 \\
-b' & 1
\end{array}
\right),\qquad
H_0^{(0,1)} =
\left(
\begin{array}{cc}
1 & -b \\
0 & z-z_0
\end{array}
\right).
\label{eq:mm_single}
\eeq
The complex parameter $z_0$ corresponds to the position of the vortex while the other parameters, $b$ and $b'$,
parameterize the internal orientation. This modulus gives rise to an internal
moduli space ${\bf C}P^1$~\cite{Eto:2005yh,Eto:2006pg}.
In fact, they are inhomogeneous coordinates for ${\bf C}P^1$ and are related by the
transition function $b = 1/b'$. This orientational modulus can easily be understood from a simple argument
related to the symmetry of the theory as we will see below.

A rigorous  way to define the
orientation of the non-Abelian vortex is to identify it with the  null eigenvector of $H_0(z)$ at the vortex position
$z=z_0$. For the moduli matrix in Eq.~(\ref{eq:mm_single}) the orientational vectors are
\beq
\vec\phi^{~(1,0)} =
\left(
\begin{array}{c}
1\\
b'
\end{array}
\right)
\quad\sim\quad
\vec \phi^{~(0,1)}
=
\left(
\begin{array}{c}
b\\
1
\end{array}
\right).\label{fundorient}
\eeq
Here ``$\sim$" stands for an identification up to complex non zero factors: $\vec \phi \sim \lambda \vec\phi$,
$\lambda \in {\bf C}^*$. One can easily find a direct relation between the parameter $b$ ($b'$) and the broken
$SU(2)_{\rm C+F}$. Since $H_0$ transforms as $H_0 \to H_0 U_{\rm F}$ under the color-flavor group, the
orientational vector $\vec \phi$ transforms as $\vec\phi \to U_{\rm F}^\dagger \vec \phi$. If we start with
$\vec\phi = (1,0)^T$, we can recover $\vec \phi = (1,b')^T$ by use of the color-flavor rotation as
\beq
\vec\phi = U_{\rm F}^\dagger \vec\phi
\quad\Leftrightarrow\quad
\left(
\begin{array}{c}
1\\
0
\end{array}
\right)
\to
\left(
\begin{array}{cc}
\alpha^* & -\beta\\
\beta^* & \alpha
\end{array}
\right)
\left(
\begin{array}{c}
1\\
0
\end{array}
\right)
\sim
\left(
\begin{array}{c}
1\\
\beta^*/\alpha^*
\end{array}
\right)
\label{eq:ori_gene}
\eeq
with $|\alpha|^2 + |\beta|^2 = 1$. Thus we identify $b'$ and $\beta^*/\alpha^*$. In what follows, we will call
two non-Abelian vortices with equal orientational vectors {\it parallel}, while when they have orthogonal orientational
vectors we will call them {\it anti-parallel}. The reader must keep in mind that vortices are always parallel in real
space. Throughout this paper, we use the words parallel and anti-parallel only referring to the internal orientational
vectors.

Generic configurations of two vortices at arbitrary positions and with arbitrary orientations are described by the moduli
matrices~\cite{Eto:2005yh,Eto:2006pg}:
\beq
H_0^{(1,1)} =
\left(
\begin{array}{cc}
z-\phi & -\eta\\
-\tilde\eta & z - \tilde\phi
\end{array}
\right),\quad
H_0^{(2,0)} =
\left(
\begin{array}{cc}
z^2-\alpha'z-\beta' & 0\\
-a'z-b' & 1
\end{array}
\right).
\label{eq:mm_gene}
\eeq
The superscripts label patches covering the moduli space. One more patch similar to $(2,0)$ is needed to complete the
full moduli space~\cite{Eto:2005yh,Eto:2006pg}. The positions of the vortices are the roots of $z_i^2 -
(\phi+\tilde\phi) z_i + \phi\tilde\phi - \eta\tilde\eta =z_i^2-\alpha'z_i-\beta'= 0$. By using translational symmetry
we can set $z_1 + z_2 = 0$ ($ \phi + \tilde \phi = \alpha'=0$) without loss of generality. The orientation vectors are
$\vec\phi^{~(1,1)}_1 = \left( \eta,\ z_1 - \phi \right)^T$ and $\vec\phi^{~(1,1)}_2 = \left( \eta,\ z_2 - \phi \right)^T$
for the $(1,1)$ patch, while they are $\vec\phi^{~(2,0)}_1 = \left( 1,\ a'z_1+b' \right)^T$ and $\vec\phi^{~(2,0)}_2 =
\left( 1,\ a' z_2+b' \right)^T $ for the $(2,0)$ patch. Overall complex factor does not have physical meaning, so that
each vector takes value on ${\bf C}P^1$. We can describe anti-parallel vortices only in the (1,1) patch when $\eta =
\tilde \eta$, because of ${\vec\phi}^{\dagger}_1 \vec\phi_2 = 0$. On the other hand, we can describe parallel vortices only
in the $(2,0)$ patch when $a'z_1+b'=a'z_2+b'$.

For convenience, let us take a special subspace where $\tilde \eta = 0$ in the $(1,1)$ patch:
\beq
H_{0 \ {\rm red}}^{(1,1)} \equiv \left(
\begin{array}{cc}
z-z_0 & -\eta\\
0 & z + z_0
\end{array}
\right), \quad z_0=z_1=-z_2=\phi=-\tilde\phi.\label{reducedpatch}
\eeq
One can always recover generic points in Eq.~(\ref{eq:mm_gene}) using  flavor rotations. The parameters $(z_0,\eta)$
and $(\beta,a',b')$ are related by the following relations: $\beta' = z_0^2, \, a' = 1/{\eta}$ and $b' = - z_0/\eta$.
The orientational vectors are then of the form
\beq
\vec\phi^{~(1,1)}_1\big|_{z=z_0} =
\left(
\begin{array}{c}
1\\
0
\end{array}
\right),\qquad
\vec\phi^{~(1,1)}_2\big|_{z=-z_0} =
\left(
\begin{array}{c}
\eta\\
-2 z_0
\end{array}
\right).
\label{eq:oris}
\eeq

When the vortices are coincident, however, the rank of the moduli matrix at the vortex position reduces. In this case
we can no longer define two independent orientations for each vortex but we can only define an overall orientation. In
fact, one can see that when $z_0 = 0$, the two orientations in Eq.~(\ref{eq:oris}) are both equal to $\vec \phi^{~(1,1)}
= (1,\ 0)^T$. We cannot really give to the parameter $\eta$ an exact physical meaning of a relative
orientation between two coincident vortices. It is better, in this case, to consider this parameter merely as an
internal degree of freedom of the composite vortex. When we take correctly into account both the parameter $\eta$ and
the global flavour rotations that we previously factorized out, we recover the full moduli space for coincident
vortices, which is $W{\bf C}P^2_{(2,1,1)} \simeq {\bf C}P^2/Z_2$~\cite{Hashimoto:2005hi,Auzzi:2005gr,Eto:2006cx}.

The definitions of the position and orientation of a single vortex can be rigorously extended to the
non-BPS case by merely replacing $H_0(z)$ with $\tilde H_0(z,\bar z)$. For configurations with several vortices,
all the flat directions that are not related to Goldstone modes or translational symmetries will disappear. It
is possible to use these definitions as constraints on the $\tilde H_0(z,\bar z)$ matrix
to fix positions and orientations. Then our formalism allows us to
study the static interactions of non-BPS configurations.

\section{Vortex interaction in the fine-tuned model}

We now concentrate on the fine-tuned model (\ref{eq:reL}). We will first calculate the masses of a special class of
non-BPS coincident vortices. Then we will derive an effective potential for coincident almost BPS vortices but with
generic value of the internal modulus parameter. Finally, we will compute an effective potential for two almost BPS
vortices at any distance and with any relative orientations.

\subsection{$(k_1,k_2)$ coincident vortices}

The minimal winding solution in the non-Abelian gauge theory is a mere embedding of the ANO solution into the
non-Abelian theory. This is obvious also from the moduli matrix view point. In fact, in the non-Abelian moduli matrix
(\ref{eq:mm_single}) we can put $b$ (or $b'$) to zero with a global flavour rotation. Henceforth we can recognize the moduli
matrix for the single ANO vortex: $H_0^{\rm ANO}(z) = z - z_0$ as the only non-trivial element of the moduli matrix.

\begin{figure}[ht]
\begin{center}
\begin{tabular}{ccc}
\includegraphics[height=5cm]{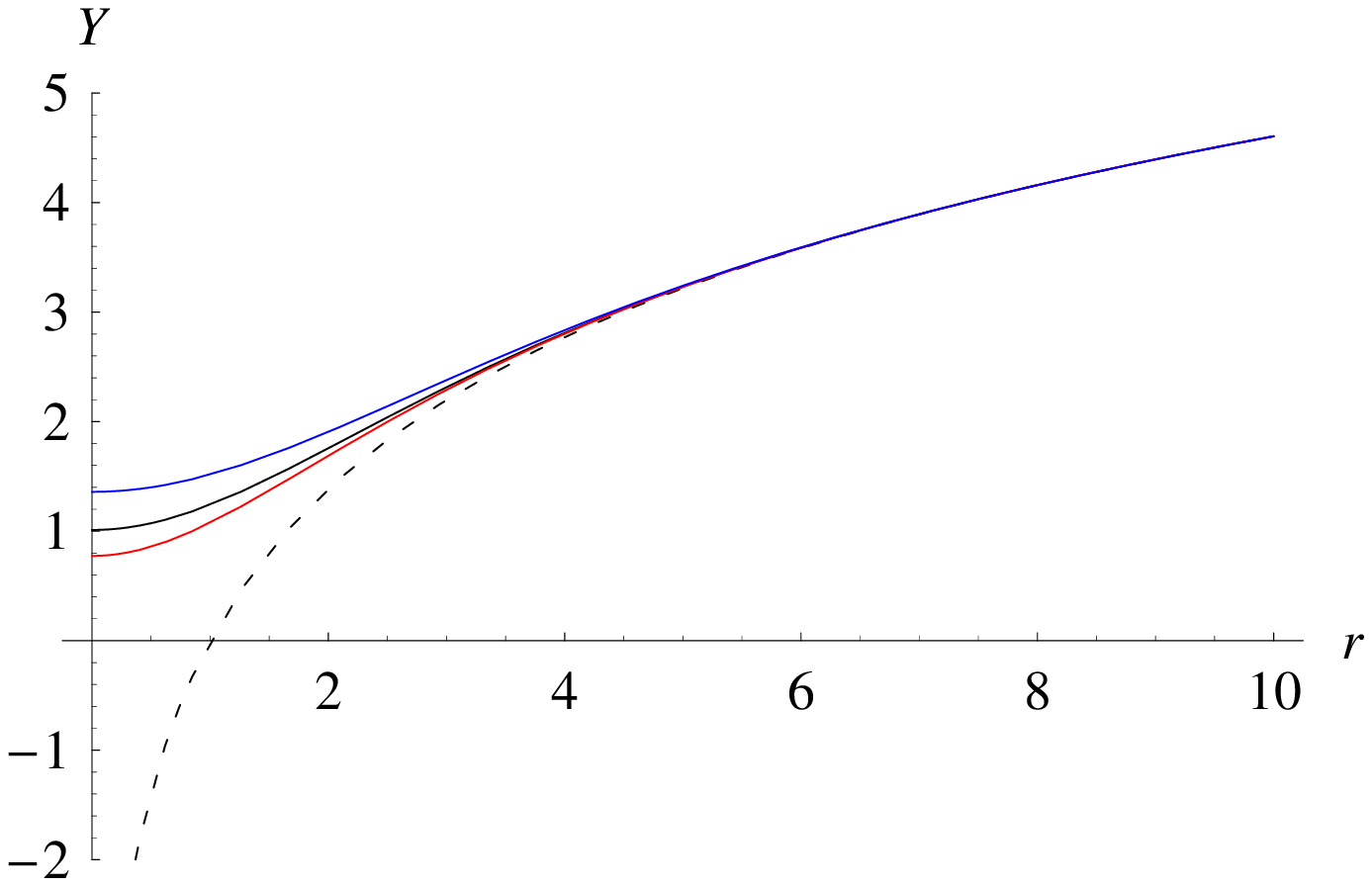}
&\qquad&
\includegraphics[height=5cm]{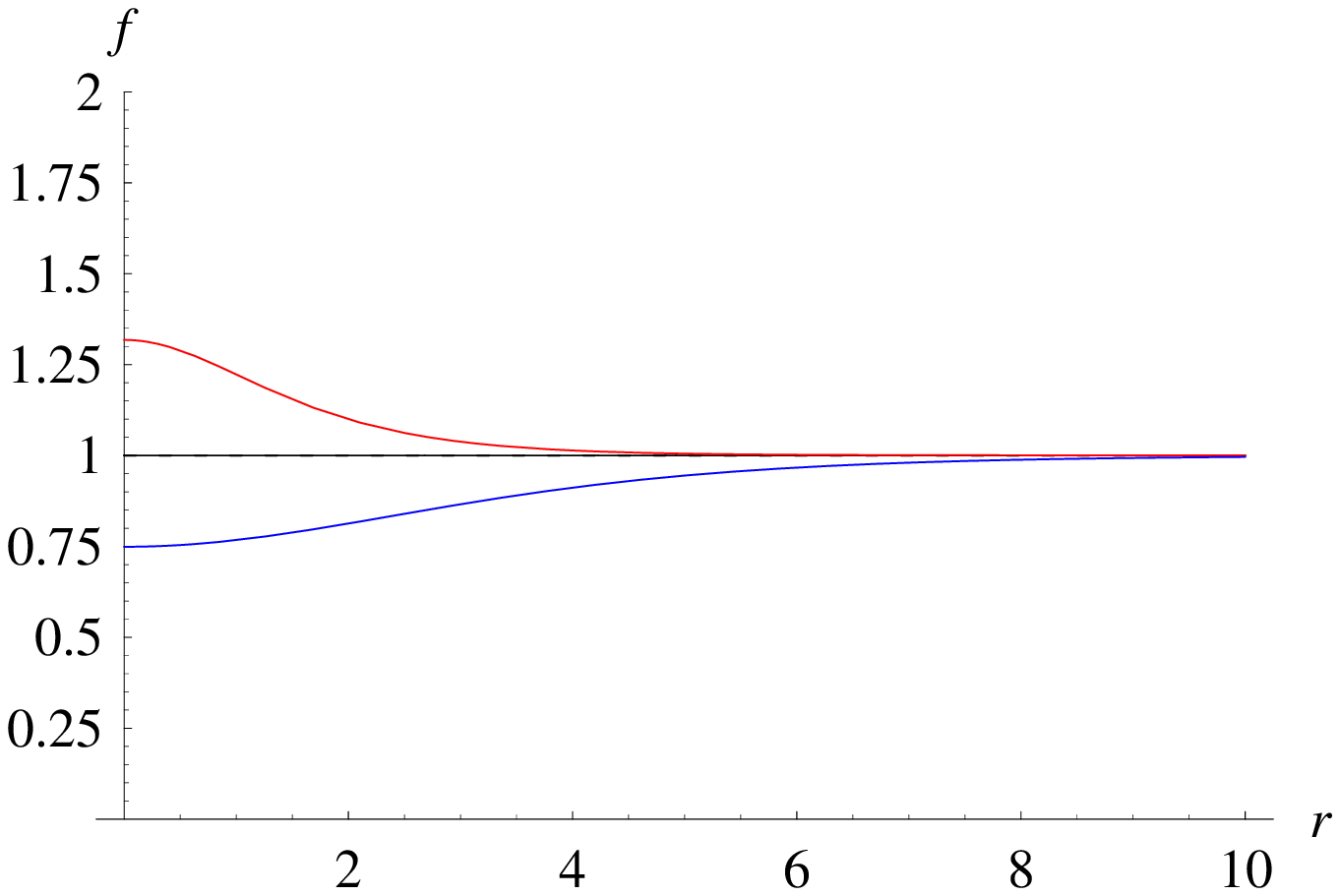}
\end{tabular}
\caption{{\small Numerical plots of $Y$ (left) and $f$ (right) for the single vortex: black for $\lambda=1$, red for
$\lambda = 1.7$, blue for $\lambda = 0.5$. The broken line is $2 \log r$.}} \label{fig:Y}
\end{center}
\end{figure}

This kind of embedding is also useful to investigate a simple non-BPS configuration.
Let us start with the moduli
matrix for a configuration of  $k$ coincident vortices. Since we have an axial symmetry
around the $k$ coincident vortices, we can make the following reasonable ansatz for $\Omega$ and $\tilde H$
\beq
\Omega^{(0,1)} =
\left(
\begin{array}{cc}
1 & 0\\
0 & w(r)
\end{array}
\right),\qquad
\tilde H^{(0,1)} =
\left(
\begin{array}{cc}
1 & 0\\
0 & f(r) z^k
\end{array}
\right).
\label{eq:mm_para}
\eeq
Note that $f(r) = 1$ means $\tilde H^{(0,1)} = H_0^{(0,1)}(z)$ which is nothing but the BPS solution. We will call the
multiple vortex which is generated by the ansatz in Eq.~(\ref{eq:mm_para}) ``$(0,k)$-vortex''. In terms of the two
fields $w(r) = e^{Y(r)}$ and $f(r)$ the equations (\ref{eq:eom_omega1}) and (\ref{eq:eom_omega2}) reduce to the
following form
\beq
&&Y'' + \frac{1}{r}Y'
- \frac{2}{f} \left( f'' + \frac{1+2k}{r}f' - Y'f' \right)
- \lambda^2 \left( 1 - r^{2k} f^2 e^{-Y} \right) = 0,
\label{eq:reduced1}\\
&&Y''' + \frac{1}{r}Y'' - \frac{1}{r^2}Y' + e^{-Y}r^{2k-1} f^2 \left( 2k - rY' \right) = 0. \label{eq:reduced2}
\eeq
The boundary conditions are
\beq
 & Y \to 2k \log r,\quad Y' \to 2 k/r,\quad f \to 0\qquad ( r \to \infty );& \\
 & Y' \to 0,\quad f' \to 0\qquad ( r \to 0 ).&
\eeq
Although it is quite impossible to solve these differential equation analytically, we can solve them numerically.
The results for the single vortex ($k=1$) are shown in Fig.~\ref{fig:Y}, where we used several different values of
$\lambda$ .

When $k \ge 2$ it is possible that the ansatz (\ref{eq:mm_para}) does not give the true solution (minimum of the energy) of
the equations of motion (\ref{eq:eom_omega1}) and (\ref{eq:eom_omega2}). This is because there could be repulsive
forces between the vortices. With ansatz (\ref{eq:mm_para}) we fix the positions of all the vortices at the origin by
hand. The reduced equations (\ref{eq:reduced1}) and (\ref{eq:reduced2}) are nevertheless still useful to investigate the
interactions between two vortices. The results are listed in Table \ref{tab:mass}.
\begin{table}[ht]
\begin{center}
\begin{tabular}{c|ccc}
$\lambda$ & $k=1$ & $k=2$ & $k=3$ \\
\hline
0.6 & 0.81305 & 1.52625 & 2.21205\\
0.7 & 0.86440 & 1.65337 & 2.42101\\
0.8 & 0.91231 & 1.77407 & 2.62115\\
0.9 & 0.95737 & 1.88936 & 2.81382\\
1   & 1.00000 & 2.00000 & 3.00000\\
1.1 & 1.04053 & 2.10655 & 3.18045\\
1.2 & 1.07922 & 2.20944 & 3.35575\\
1.3 & 1.11626 & 2.30905 & 3.52639\\
1.4 & 1.15182 & 2.40566 & 3.69276\\
\end{tabular}
\caption{{\small Numerical value for the masses of coincident vortices.}} \label{tab:mass}
\end{center}
\end{table}
For $\lambda = 1$, the masses are identical to integer values, up to $10^{-5}$ order, which are nothing but the winding
number of the vortices. Furthermore, our numerical results for generic $\lambda$ are in perfect agreement with the
numerical value for ANO vortices obtained about 30 years ago by Jacobs and Rebbi~\cite{Jacobs:1978ch}. As mentioned, this
happens because the $(0,k)$-vortex is obtained by embedding of the $k$ ANO vortices.

There is another type of composite configuration which can easily be analyzed numerically. These configurations are
generated by the following ansatz for $\Omega$ and $\tilde H$
\beq
\Omega^{(1,1)} =
\left(
\begin{array}{cc}
w_1(r) & 0\\
0 & w_2(r)
\end{array}
\right),\qquad
\tilde H^{(1,1)} =
\left(
\begin{array}{cc}
f_1(r) z^{k_1} & 0\\
0 & f_2(r) z^{k_2}
\end{array}
\right).
\label{eq:mm_anti_para}
\eeq
This ansatz corresponds to a configuration with $k_1$ composite vortices which wind in the first diagonal $U(1)$ subgroup
of $U(2)$ and with $k_2$ coincident vortices that wind the second diagonal $U(1)$ subgroup. The two sets of vortices
can be considered each as embedded ANO vortices for the two decoupled Abelian subgroups.
We refer to these decoupled non-Abelian vortices as a ``$(k_1,k_2)$-vortex''. The mass of a $(k_1,k_2)$-vortex is thus the
sum of the mass of the $(k_1,0)$-vortex and that of the $(0,k_2)$-vortex. For example, the mass of $(1,1)$-vortex is
double of the mass of the $(0,1)$-vortex listed in the first column of Table \ref{tab:mass}.
As in the previous case, we get the  minima of the energy under the constraint that the vortices are coincident.

\begin{figure}[ht]
\begin{center}
\includegraphics[width=14cm]{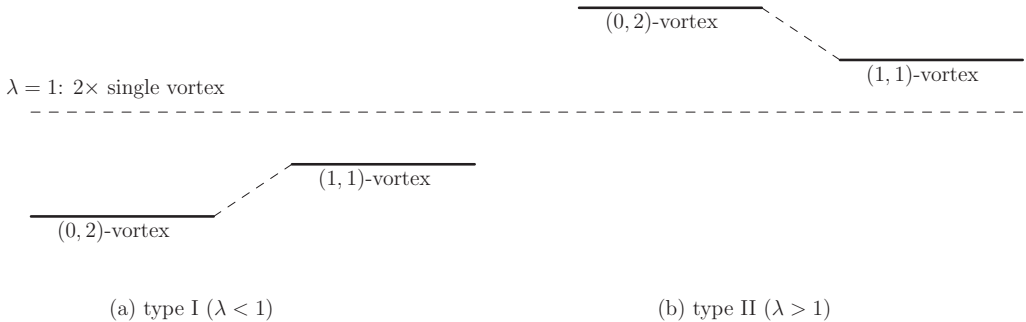}
\caption{{\small Spectrum of the $(0,2)$ and $(1,1)$ coincident vortices.}}
\label{fig:spectrum}
\end{center}
\end{figure}
Because our fine-tuned non-Abelian model is a simple extension of the Abelian-Higgs model, we expect similar behavior
for the interactions. Actually we have only one parameter $\lambda$. Thus we will call the non-Abelian vortices for
$\lambda < 1$ type I, while they will be called type II for $\lambda>1$. From Fig.~\ref{fig:spectrum}, in which is summarized
the relevant data of Table \ref{tab:mass}, we can argue which kind of interaction appears between two non-Abelian
vortices. In the type I case, the $(0,2)$-vortex is energetically preferred to the $(1,1)$-vortex, while in type II case the
$(1,1)$-vortex is preferred. If the two vortices are separated sufficiently, we can ignore any interaction between them.
Regardless of their orientations, the mass of two well separated vortices is twice that of the single vortex. This mass
is equal to the mass of the $(1,1)$-vortex. Furthermore, it seems that the two separated anti-parallel vortices do not
interact, and the energy does not depend on the relative distance.
For the type I case, Fig.~\ref{fig:spectrum} suggests that
the configuration with $(2,0)$-vortices is the true minimum of the system. This means that attractive forces appears
between vortices with different orientations. An attractive force also works in the internal space, which aligns the
orientations.
In the type II case, it seems that we do not have an isolated minimum of the energy. In fact, all the anti-parallel
configurations with arbitrary distance have the same value of the energy. In the following sections we will confirm
the picture we have outlined here.

\subsection{Effective potential for coincident vortices \label{sect:eff_coinc}}

The dynamics of BPS solitons can be investigated by the so-called moduli approximation~\cite{Manton:1981mp}. The
effective action is a massless non-linear sigma model whose target space is  the moduli space. The sigma model is
obtained by plugging a BPS solution into the original Lagrangian and promoting the moduli parameters to massless
fields, then picking up quadratic terms in the derivatives with respect to the vortex world-volume coordinates
\beq
L = \int dx^1dx^2\ {\cal L}\left[H_{\rm sol}(\varphi_i(t,x^3)),W^\mu_{\rm sol}(\varphi_i(t,x^3))\right]_{\lambda=1},
\eeq
where $\varphi_i$ represents the set of moduli parameters $(\eta,\tilde \eta, \phi, \tilde \phi)$ or $(\alpha',
\beta', a',b')$ contained in the moduli matrix (\ref{eq:mm_gene}).

If the coupling constant $\lambda$ is close to the BPS limit $\lambda = 1$, we can still use the moduli
approximation, to investigate dynamics of the non-BPS non-Abelian vortices by adding a potential of order $|1-\lambda^2|
\ll 1$ to the massless sigma model
\beq
L = \int dx^1dx^2\ {\cal L}\left[H_{\rm sol}(\varphi_i(t,x^3)),W^\mu_{\rm sol}(\varphi_i(t,x^3))\right]_{\lambda=1} -
V(\varphi_i). \label{eq:eff_lag}
\eeq
We shall now calculate the effective potential $V(\varphi_i)$ using the method suggested by Hindmarsh, who calculated
this effective potential for non-BPS semilocal vortex in the Abelian-Higgs model \cite{Hindmarsh:1992yy}.

First we write the Lagrangian (\ref{eq:reL}) in the following way
\beq
\tilde {\cal L} = \tilde {\cal L}_{\rm BPS}+\frac{(\lambda^2 - 1)}{4} \left( {\bf 1}_{N} - HH^\dagger \right)^2.
\label{eq:Lag_Hind}
\eeq
We get non-BPS corrections of order $O(\lambda^2 - 1)$ by putting BPS solutions
into Eq.~(\ref{eq:Lag_Hind}). This is
because the first term is minimized by the BPS solution, while the second one is already a term of order $O(\lambda^2 -
1)$. The energy functional thus takes the following form
\beq
{\cal E}=\frac{E}{2 \pi v^2} = 2   + \frac{(\lambda^2 - 1)}{8 \pi}  \int dx^1dx^2 \ \Tr \left({\bf 1} - H_{\rm
BPS}(\varphi_i)H_{\rm BPS}^\dagger(\varphi_i) \right)^2\label{eff gen}
\eeq
where $H_{\rm BPS}(\varphi_i)$ stands for the BPS solution  generated by the moduli matrices in
Eq.~(\ref{eq:mm_gene}). The first term corresponds to the mass of two BPS vortices and the second term is the deviation
from the BPS solutions which is nothing but the effective potential we want.

In this section we consider the effective potential on the moduli space of coincident vortices. To this end, it suffices
to consider only the following matrices
\beq
H_0^{(1,1)} = \left(
\begin{array}{cc}
z & -\eta\\
0 & z
\end{array}
\right),\quad H_0^{(2,0)} = \left(
\begin{array}{cc}
z^2 & 0\\
-a'z & 1
\end{array}
\right). \label{eq:mm_coinc}
\eeq
The parameters $\eta$ and $a'$ are related by $\eta=1/a'$. The effective potential on the moduli space for two
coincident vortices is thus \footnote{ The potential depends only on $|\eta|$ because the phase of $\eta$ can be
absorbed by the flavor symmetry. The same holds for the coordinate $a'$ }
\beq
\frac{V(|\eta|)}{2 \pi v^2} \equiv \frac{(\lambda^2 - 1)}{8 \pi}  \int dx^1dx^2 \ \Tr \left({\bf 1} - H_{\rm
BPS}(|\eta|)H_{\rm BPS}^\dagger(|\eta|) \right)^2\equiv(\lambda^2 - 1) {\cal V}(|\eta|), \label{eq:eff_V}
\eeq
where we have defined a reduced effective potential ${\cal V}$ which is independent of $\lambda$. To evaluate this
effective potential, we need to solve the BPS equations for a composite state of two non-Abelian vortices with
an intermediate value of $\eta$. Such numerical solutions found in \cite{Auzzi:2005gr}. We propose here another
reliable technique for the numerics, which needs much less algebraic efforts.
\begin{figure}[ht]
\begin{center}
\begin{tabular}{ccc}
\includegraphics[width=8cm]{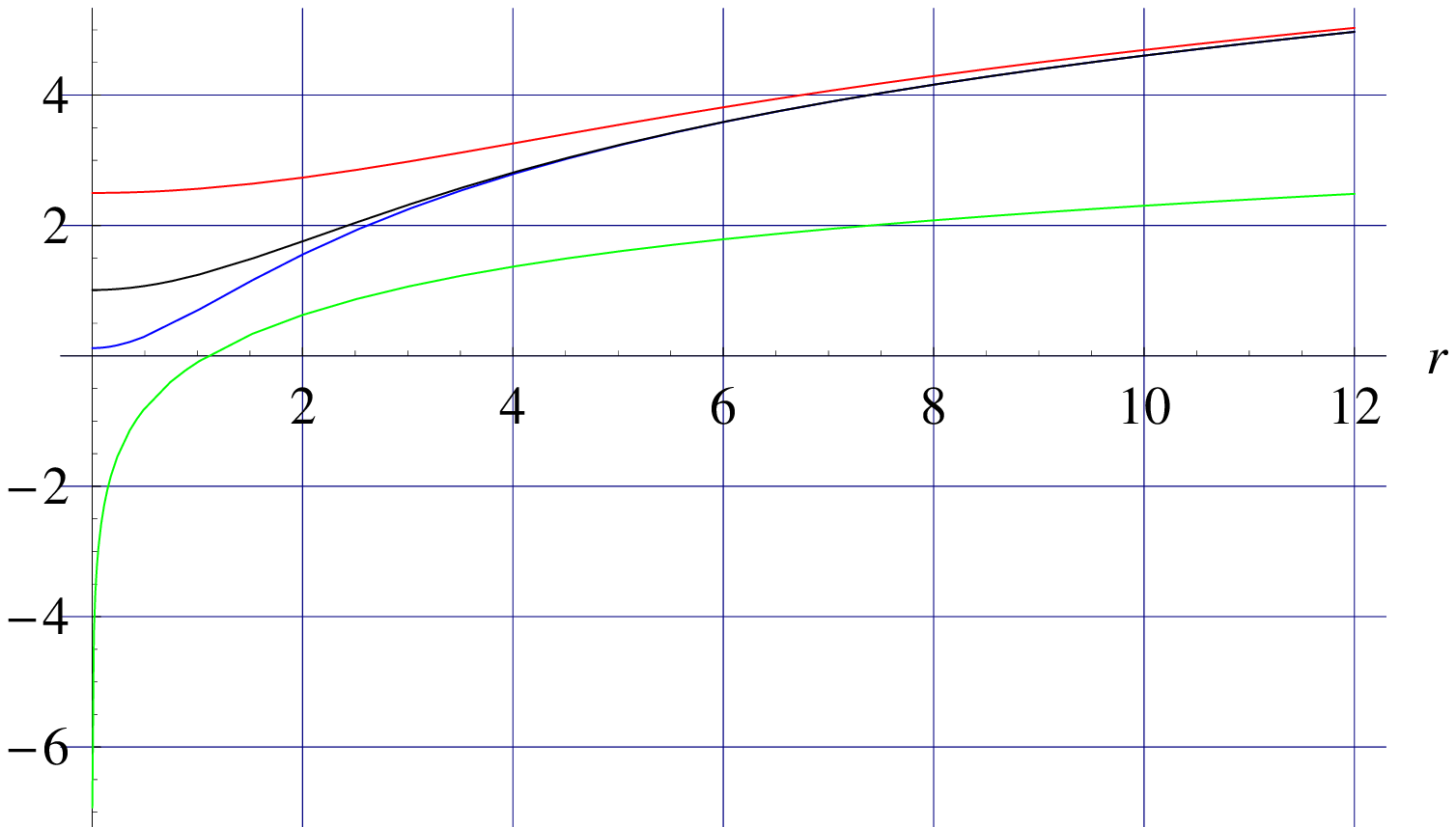}
& \qquad &
\includegraphics[width=8cm]{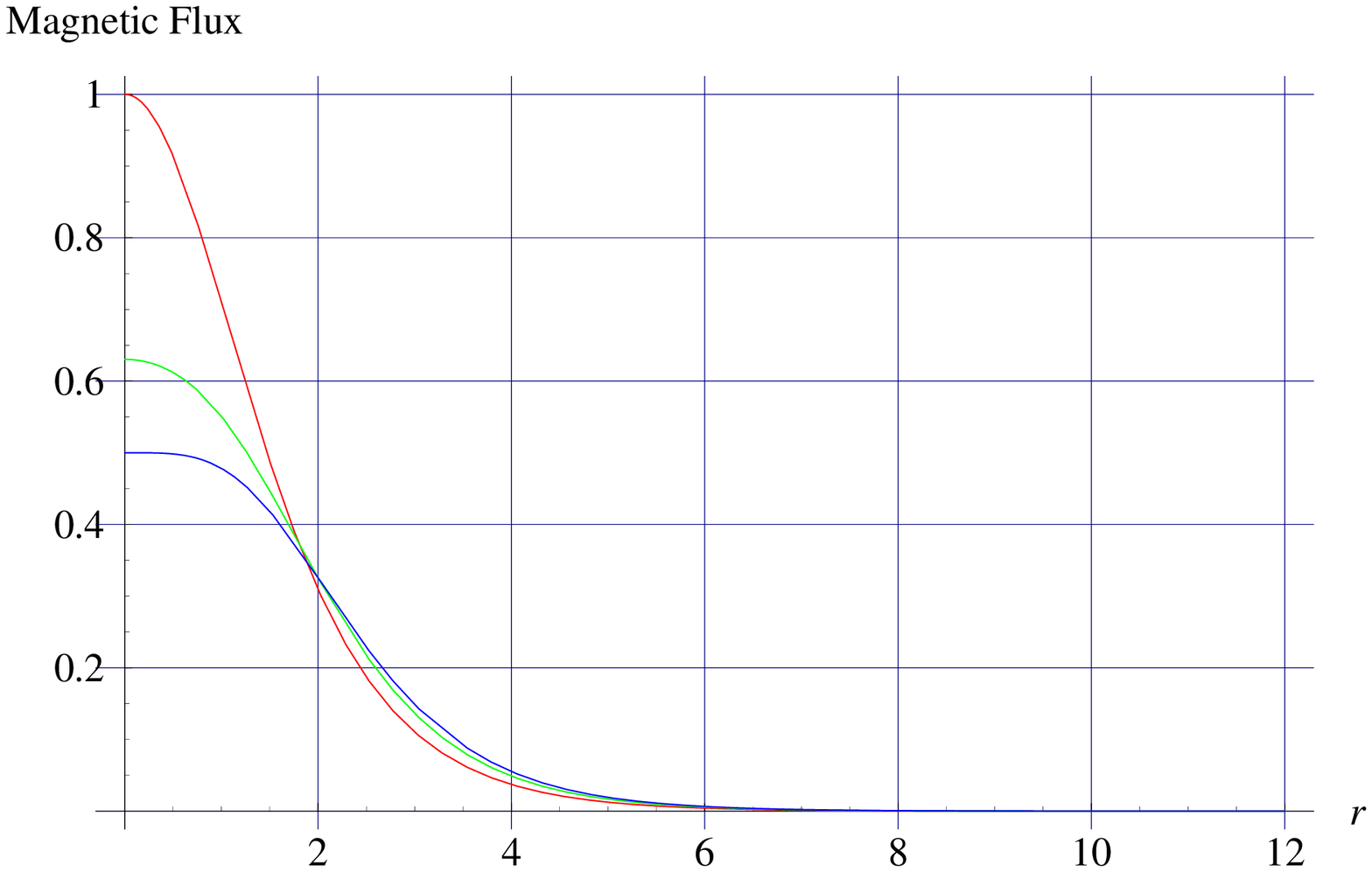}
\end{tabular}
\caption{{\small Left: Numerical plots of $Y_1$ (red), $Y_2$ (green) and $Y_3$ (blue) for $\eta=3$. The black line is
$Y_1=Y_3$ for $\eta=0$. Right: Magnetic flux $\Tr F_{12}$ with $\eta=0$ (red), $\eta=3$ (green) and $\eta=\infty\ (a'=0)$
(blue).}} \label{fig:mflux}
\end{center}
\end{figure}
In the moduli matrix formalism, what we should solve is only the master equation (\ref{eq:master}) for $\Omega$ with
$\lambda = 1$ and $\tilde H(z, \bar z) = H_0(z)$.
Because of the  axial symmetry of the composite vortex and the
boundary condition at infinity:
\beq \Omega  \rightarrow H_0(z)H_0^\dagger(\bar z), \eeq
 we can make a simple ansatz for
$\Omega$. For example in the patch $(1,1)$ we can write
\beq
\Omega^{(1,1)} = \left(
\begin{array}{cc}
w_1(r) & -\eta e^{-i\theta}w_2(r) \\
-\eta e^{i\theta}w_2(r) & w_3(r)
\end{array}
\right).
\label{eq:omega_11}
\eeq
The advantage of the moduli matrix formalism is that only three functions $w_i(r)$ are needed and the formalism itself is gauge
invariant. Plugging the ansatz into Eq.~(\ref{eq:master}), after some algebra we get the following differential
equations
\beq
Y_1''+\frac{1}{r}Y_1' + \frac{|\eta|^2}{r^2} \frac{(1+rY_1'-rY_2')^2}{|\eta|^2 - e^{Y_1+Y_3-2Y_2}} = 1 - e^{-Y_1}(r^2 +
|\eta|^2);
\label{eq:eq_Y1}\\
Y_2''+\frac{1}{r}Y_2' - \frac{1}{r^2} \frac{(1 + rY_1' - rY_2')(1+r Y_2' - rY_3')}{1 - |\eta|^2 e^{-Y_1-Y_3+2Y_2}} = 1
- e^{-Y_2}r;
\label{eq:eq_Y2}\\
Y_3''+\frac{1}{r}Y_3' + \frac{|\eta|^2}{r^2} \frac{(1+rY_2'-rY_3')^2}{|\eta|^2 - e^{Y_1+Y_3-2Y_2}} = 1 - e^{-Y_3}r^2,
\label{eq:eq_Y3}
\eeq
where we have redefined the fields as $w_i(r) = e^{Y_i(r)}$ with $i=1,2,3$. We solve numerically these differential
equations using a simple relaxation method, see Fig.\ref{fig:mflux}.

\begin{figure}[ht]
\begin{center}
\begin{tabular}{ccc}
\includegraphics[width=7cm]{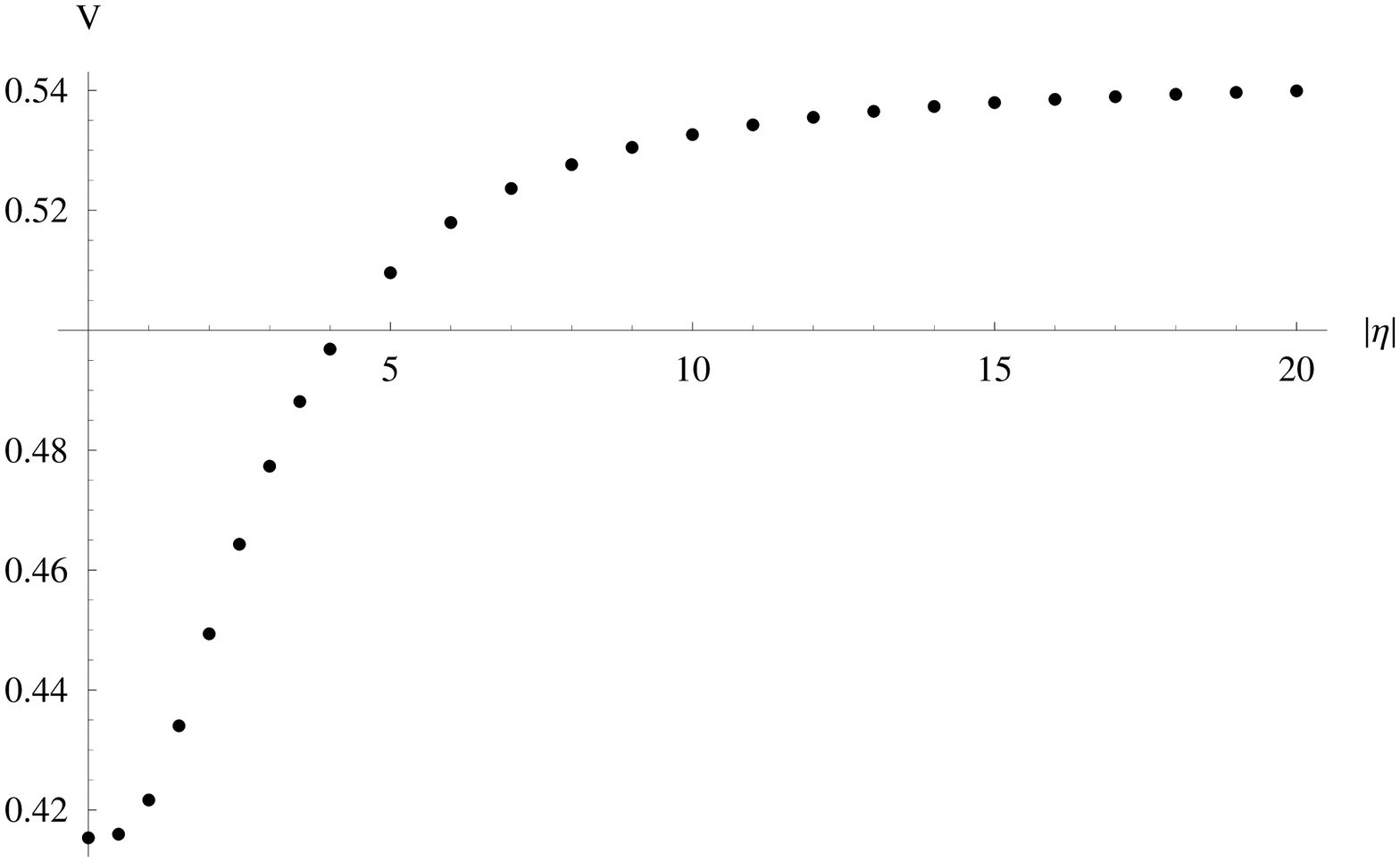}
& \qquad &
\includegraphics[width=7cm]{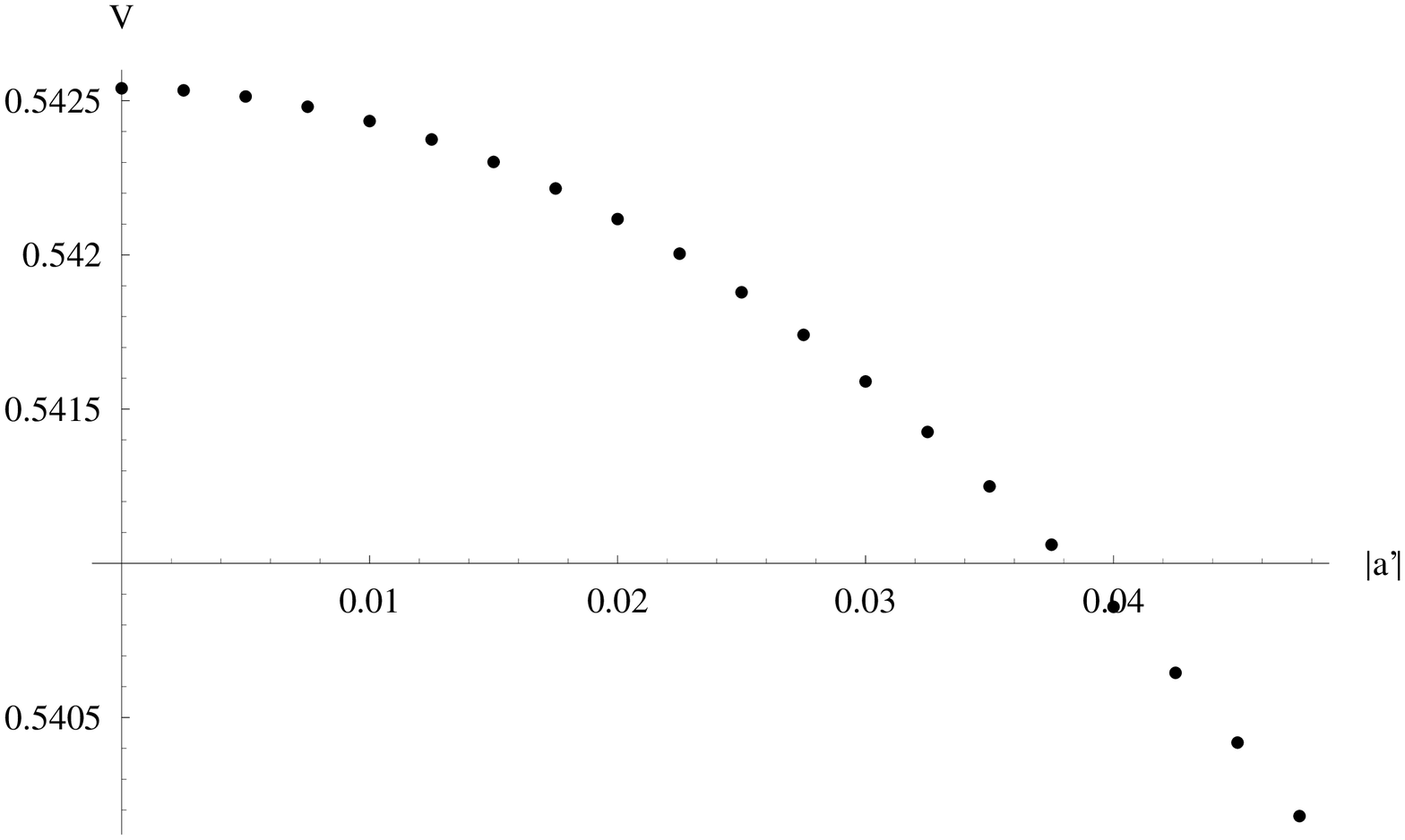}
\end{tabular}
\caption{{\small Numerical plots of the effective reduced potential ${\cal V}(|\eta|)$. In the second plot
$a'=1/\eta$.}} \label{fig:num_V}
\end{center}
\end{figure}

The effective potential can be obtained by plugging numerical solutions into Eq.~(\ref{eq:eff_V}). The result is shown
in Fig.~\ref{fig:num_V}.
When we consider configurations with big values of $|\eta|$, it is better to switch to the other patch and use the
variable $a'$. We can still make a similar ansatz for $\Omega$, leading to simple differential equations like those in
Eqs.~(\ref{eq:eq_Y1})$\sim$(\ref{eq:eq_Y3}). In the left of Fig.~\ref{fig:num_V} we show a numerical plot of the reduced
effective potential ${\cal V}$ from $|\eta|=0$ ($|a'|=\infty$) to $|\eta|=20$. In the right of Fig.~\ref{fig:num_V} we
show another plot from $|a'| = 0$ ($|\eta|=\infty$) to $|a'|=1/20$. To give an estimate of the range of validity of
our approximation we can compare the results obtained in this section with the numerical integrations obtained for
$(2,0)$-vortices and also $(1,1)$-vortices. The comparison is made in Table \ref{tab:validity} from which we can argue
that the effective potential approximation gives result with an accuracy around $10\%$ for the range of values $0.7 <
\lambda < 1.15$.
\begin{table}[ht]
\begin{center}
\begin{tabular}{c|cc|cc}
$\lambda$ & $(2,0)_{\rm num}$ & $(2,0)_{\rm eff}$ & $(1,1)_{\rm num}$ & $(1,1)_{\rm eff}$\\
\hline
0.6  & 1.52625 & 1.65280 & 1.62611 & 1.73440\\
0.7  & 1.65337 & 1.72332 & 1.72880 & 1.78835\\
0.9  & 1.88936 & 1.89692 & 1.91473 & 1.92115\\
0.95 & 1.94523 & 1.94711 & 1.95793 & 1.95954\\
1    & 2.00000 & 2.00000 & 2.00000 & 2.00000\\
1.05 & 2.05376 & 2.05561 & 2.04102 & 2.04254\\
1.1  & 2.10655 & 2.11393 & 2.08106 & 2.08715\\
1.15 & 2.15843 & 2.17496 & 2.12018 & 2.13384 \\
1.2  & 2.20944 & 2.23870 & 2.15843 & 2.18260\\
\end{tabular}
\caption{{\small Numerical value for the masses of coincident vortices.
$(2,0)_{\rm num}$ is for the numerical results while $(2,0)_{\rm eff}$ is for
our approximation using the effective potential.}} \label{tab:validity}
\end{center}
\end{table}

In the type II case ($(\lambda^2-1)>0$) the effective
potential has the same qualitative behavior as showed in the figure. As we expected, it has a minimum at $|\eta|=0$. This
matches the previous result that the $(1,1)$-vortex is energetically preferred to the $(2,0)$-vortex. In the
type I case ($(\lambda^2-1)<0$) the shape of the effective potential can be obtained just by flipping the overall sign of
the effective potential of the type II case. Then the effective potential always takes a negative value, which is consistent
with the fact that the masses of the type I vortices are less than that of the BPS vortices, see the middle column in
Table \ref{tab:mass}. Contrary to the type II case, the type I potential has a minimum at $|a'|=0$ ($|\eta| = \infty$).
This means that the $(2,0)$-vortex is preferred with respect to the $(1,1)$ vortex.

\subsection{Interaction  at generic vortex separation}

In this  subsection we go on investigating the interactions of non-Abelian vortices in the $U(2)$ gauge group at generic
distances. As in Sect.~\ref{sect:eff_coinc}
we will use the moduli space approximation, considering only small deviations from the BPS
case. The generic configurations are described  by the moduli matrices in Eq.~(\ref{eq:mm_gene}). We will consider here
only the reduced $(1,1)$ patch defined in Eq.~(\ref{reducedpatch}). By putting the two vortices on the real axis we can
reduce $z_0$ to a real parameter $d$. Furthermore, by the flavor symmetry, we can freely put $\tilde \eta = 0$ and suppress
the phase of $\eta$. The relevant configurations will be described by the following moduli matrix:
\beq
H_{0 \ {\rm red}}^{(1,1)} = \left(
\begin{array}{cc}
z-d & -\eta\\
0 & z+d
\end{array}
\right),
\eeq
where $2d$ is the relative distance and $\eta$ the relative orientation.

\begin{figure}[ht]
\begin{center}
\begin{tabular}{cccc}
\includegraphics[width=7.5cm]{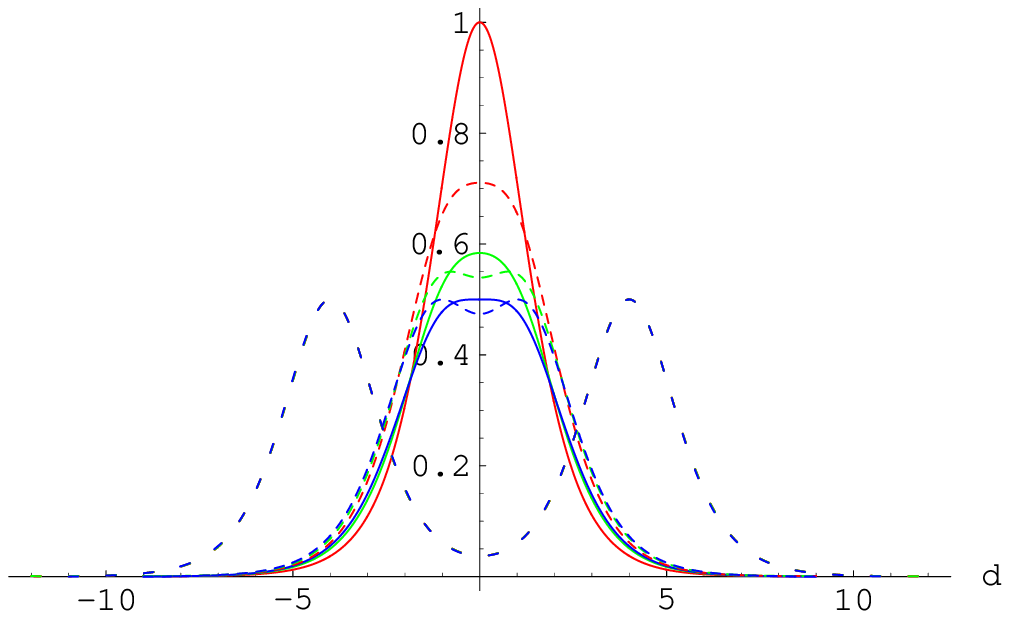} &&&
\includegraphics[width=7.5cm]{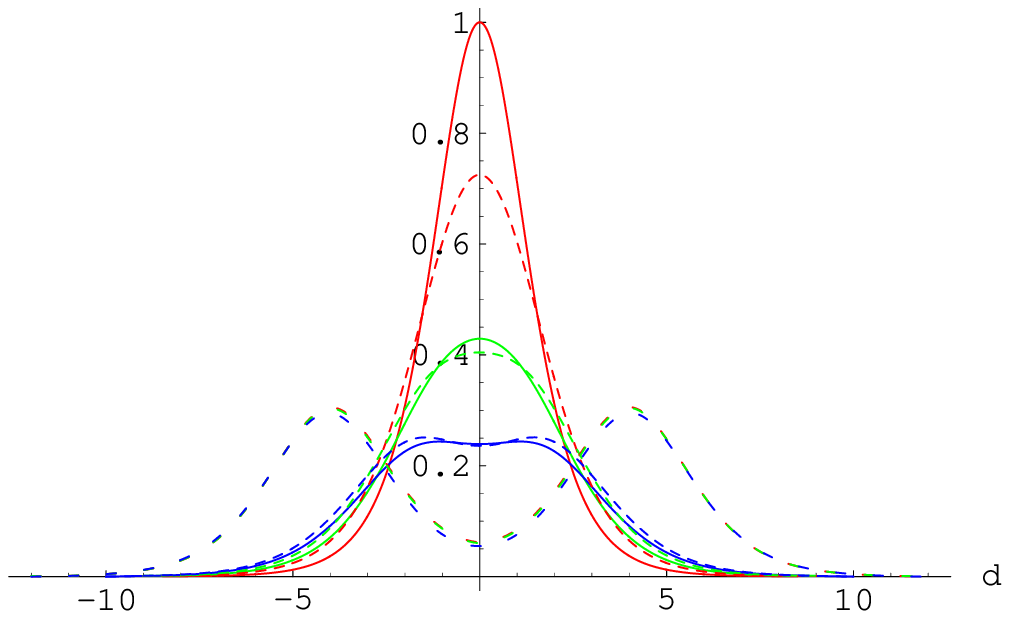}
\end{tabular}
\caption{{\small The magnetic flux $\Tr F_{12}$ of a configuration of 2 BPS vortices. The figure shows slices
including the centers of the vortex. (red, green, blue) correspond to $\eta=(0,4,\infty)$ while (solid, small broken,
wide broken) lines to $d=(0,1,4)$. The left panel shows the fine-tuned model with $e=g$ and the right shows the
model with $e=2g$.}} \label{fig:numerics}
\end{center}
\end{figure}
Now let us study the effective potential as function of $\eta$ and $d$.
We first need the numerical solution to the BPS master
equation for two vortices with any relative distance and orientation. Unlike the computation for the coincident
vortices, we do not have an axial symmetry. We can no longer reduce the problem to one spatial dimension by making an
appropriate ansatz. Nevertheless the moduli matrix formalism is a powerful tool also for the numerical
calculations. The master equation is a 2 by 2 hermitian  matrix, so it includes four real 2nd order partial differential
equations. Despite the great complexity of this system of coupled equations, the  relaxation method is very effective
to solve the problem. We show several numerical solutions in Fig.~\ref{fig:numerics}.
As before, once we get the numerical solution to the BPS equations, the effective potential is
obtained by plugging them into Eq.~(\ref{eff gen}).

\begin{figure}[ht]
\begin{center}
\includegraphics[width=10cm]{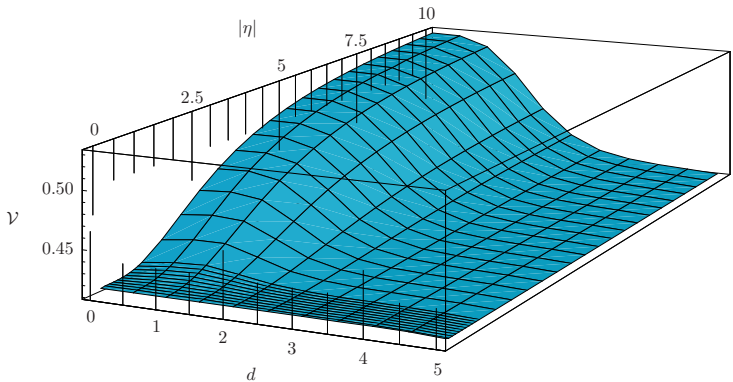}
\caption{{\small Numerical plot of the reduced effective potential ${\cal V}(\eta,d)$.}} \label{fig:effv_sep}
\end{center}
\end{figure}

The numerical plot of the reduced effective potential ${\cal V}$
is shown in Fig.~\ref{fig:effv_sep}. The effective potential for
the type II case has the same shape, up to a small positive factor ($\lambda^2-1$).
The potential forms a hill whose top is at $(d,|\eta|)= (0,\infty)$. It clearly shows that two vortices feel
repulsive forces, in both the real and internal space, for every distance and relative orientation. The minima of the
potential has a flat direction along the $d$-axis where the orientations are anti-parallel $(\eta=0)$ and along the $\eta$
axis at infinite distance $(d=\infty)$.
Therefore the anti-parallel vortices
do not interact.

In the type I case ($\lambda < 1$) the effective potential is upside-down of that of the type II case. There is unique
minimum of the potential at $(d,|\eta|)= (0,\infty)$. This means that attractive force works not only for the distance
in real space but also among the internal orientations. Configurations with anti-parallel orientations do not
interact, but these configurations represent unstable points of equilibrium. Type I vortices always stick together.

\section{Vortices with generic couplings}

In this section we will shift from the fine-tuned $U(N)$ model (\ref{eq:Lag_NAH}) to the more general model
defined in Eqs.~(\ref{flum}) and (\ref{eq:flum_pot}). This will lead to more complicated algebra, but
we will also clarify interactions with different qualitative behaviors. There are 5 parameters
$(g,e,\lambda_g,\lambda_e,v)$ in the model, but we can reduce their number with the following rescaling
\beq
H \to v H,\quad W_\mu \to e v W_\mu,\quad x_\mu \to \frac{x_\mu}{e v}.
\eeq
Then the Lagrangian (\ref{flum}) is expressed as follows
\beq
\tilde {\cal L} = \Tr \left[ - \frac{1}{2\gamma^2} \hat F_{\mu\nu} \hat F^{\mu\nu} - \frac{1}{2}
f_{\mu\nu} f^{\mu\nu} + \D_\mu H (\D^\mu H)^\dagger \right] - \frac{\gamma^2\lambda_g^2}{4} \Tr \hat X^2 -
\frac{\lambda_e^2}{4} \Tr \left(X^0T^0 - {\bf 1}_N\right)^2, \label{flam}
\eeq
where $\tilde {\cal L} = {\cal L}/{e^2v^4}$ and we have introduced the ratio of the two gauge couplings
\beq
\gamma \equiv \frac{g}{e}.
\eeq
We have three effective couplings $\gamma,\lambda_e,\lambda_g$ and the rescaled masses of particles are
\beq
M_{U(1)} = 1,\quad M_{SU(N)} = \gamma,\quad M_{\rm s} = \lambda_e,\quad M_{\rm ad} = \gamma \lambda_g.\label{masses}
\eeq
 The Lagrangian can be thought of as the bosonic part of a supersymmetric model only when both the parameters $\lambda_e$ and
$\lambda_g$ are unity. In the BPS case we can easily find the BPS equations
\beq
\bar\D H = 0,\qquad \hat F_{12} = \frac{\gamma^2}{2}\hat X,\qquad F_{12}^0T^0 = \frac{1}{2}\left(X^0T^0
- {\bf 1}_N\right). \label{eq:gene_BPS}
\eeq
The last two equations can be rewritten in a compact way
\beq
F_{12}
= \frac{1}{2} \left( X -  {\bf 1}_N\right) + \frac{\gamma^2-1}{2}\hat X. \label{eq:2ndBPS}
\eeq
The first BPS equation in Eq.~(\ref{eq:gene_BPS}) can be solved using the moduli matrix in the usual way
\beq
H = S^{-1} (z,\bar z) H_0(z), \qquad \bar W = -iS^{-1}\bar\p S,
\eeq
where $S$ is a $GL(N,{\bf C})$ matrix. We would like to stress that this solution does not depend on $\gamma$ so that the
moduli space of the BPS vortices is the same as that of the well investigated vortices in the equal gauge coupling theory
$g=e$.\footnote{The moduli space is the same from the topological point of view, while the metric will be different.}
The Eq.~(\ref{eq:2ndBPS}) can be rewritten in a gauge invariant fashion as
\beq
\bar\p \left(\Omega \p \Omega^{-1} \right) = \frac{1}{4} \left( \Omega_0 \Omega^{-1} - {\bf 1}_N\right) +
\frac{\gamma^2 - 1}{4} \left( \Omega_0 \Omega^{-1} - \frac{\Tr \left(\Omega_0 \Omega^{-1}\right)}{N} {\bf 1}_N \right)
\label{eq:gene_master}
\eeq
where $\Omega = SS^\dagger$ is same as before and $\Omega_0 \equiv H_0H_0^\dagger$.

Now we are ready to investigate interactions between two almost BPS vortices by using same strategy as we used in
Sec.~4. An effective action of the moduli dynamics for appropriately small $|1-\lambda_{e,g}^2| \ll 1$
is obtained by plugging BPS solutions into the action. Then we get
\beq
\frac{V(\eta,d;\gamma,\lambda_e,\lambda_g)}{2 \pi v^2} &=&  \int d \tilde x^2 \left(
\frac{\gamma^2(\lambda_g^2-1)}{8 \pi} \Tr \hat X^2 + \frac{\lambda_e^2-1}{8 \pi} \Tr \left(X^0T^0 - {\bf 1}_N\right)^2
\right)\nonumber\\
&=& \frac{1}{2 \pi} \int d\tilde x^2\ \Tr\left[ \frac{\lambda_g^2-1}{\gamma^2} (\hat F_{12})^2 + (\lambda_e^2-1)
(F_{12}^0T^0)^2\right],
\eeq
where we have used the BPS equations in the second line. Let us define the Abelian and the non-Abelian potentials as
\beq
{\cal V}_{e}(\eta,d;\gamma) = \int d\tilde x^2\ \Tr (F_{12}^0T^0)^2,\qquad {\cal V}_{g}(\eta,d;\gamma) =  \int d\tilde
x^2\ \Tr (\hat F_{12})^2. \label{potentials}
\eeq
The true potential is a linear combination of them
\beq
V(\eta,d;\gamma,\lambda_e,\lambda_g) = (\lambda_e^2 - 1) {\cal V}_e(\eta,d;\gamma) + \frac{\lambda_g^2
-1}{\gamma^2} {\cal V}_g(\eta,d;\gamma). \label{eq:gene_effpot}
\eeq
Notice that ${\cal V}_{e,g}$ is determined by the BPS solutions, so it does not depend on $\lambda_{e,g}$.

\subsection{Equal gauge coupling $\gamma=1$ revisited}

\begin{figure}[ht]
\begin{center}
\begin{tabular}{ccccc}
\includegraphics[width=5cm]{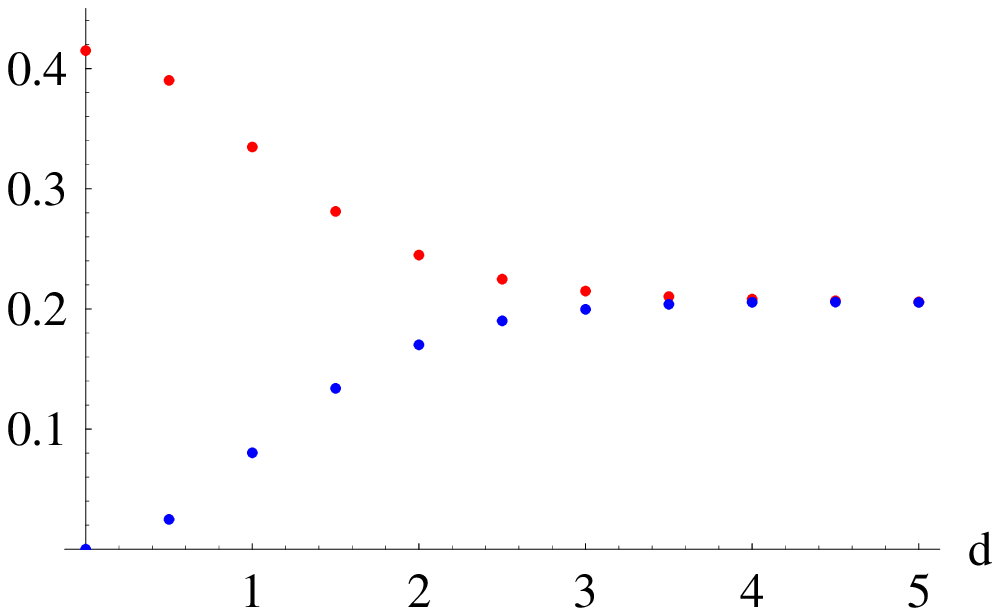}
&&
\includegraphics[width=5cm]{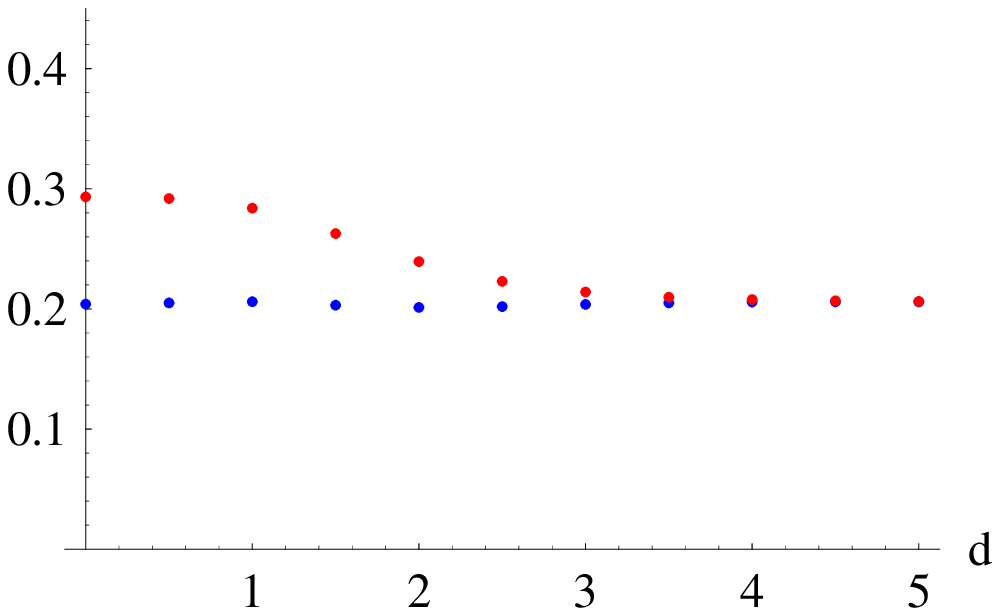}
&&
\includegraphics[width=5cm]{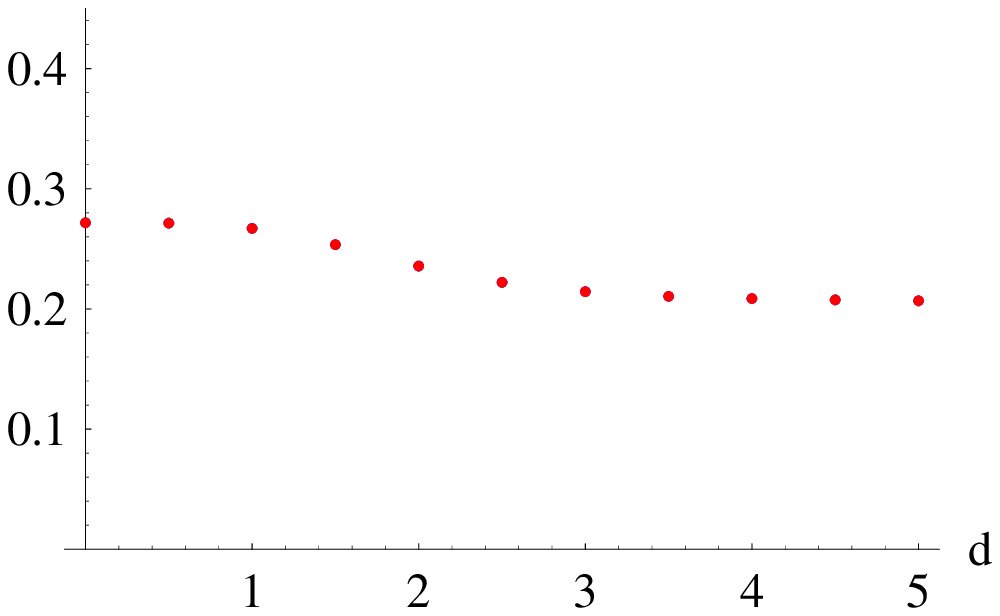}\\
anti-parallel && intermediate && parallel\\
\end{tabular}
\caption{{\small Effective potential vs. distance of vortices $d$. Red dots denote ${\cal V}_e$
while blue denote ${\cal V}_g$.
All plots are for $\gamma=1$. The left figure shows the two anti-parallel vortices $\eta=0$, the middle shows
$\eta=4.1$ and the right shows the two parallel vortices $\eta=\infty$.}} \label{fig:effv_e=g}
\ \\
\begin{tabular}{ccccc}
\includegraphics[width=5cm]{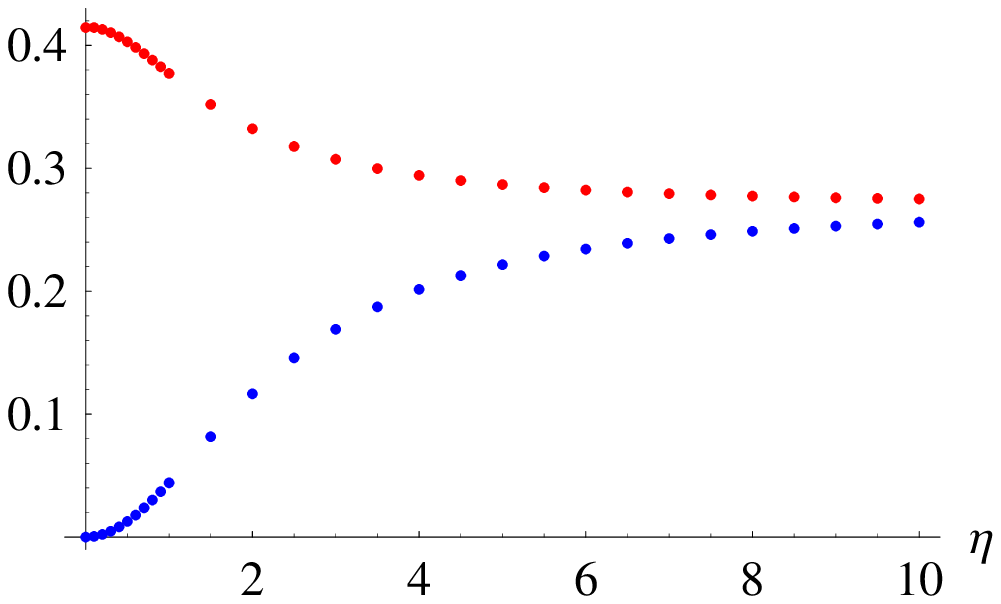}
&&
\includegraphics[width=5cm]{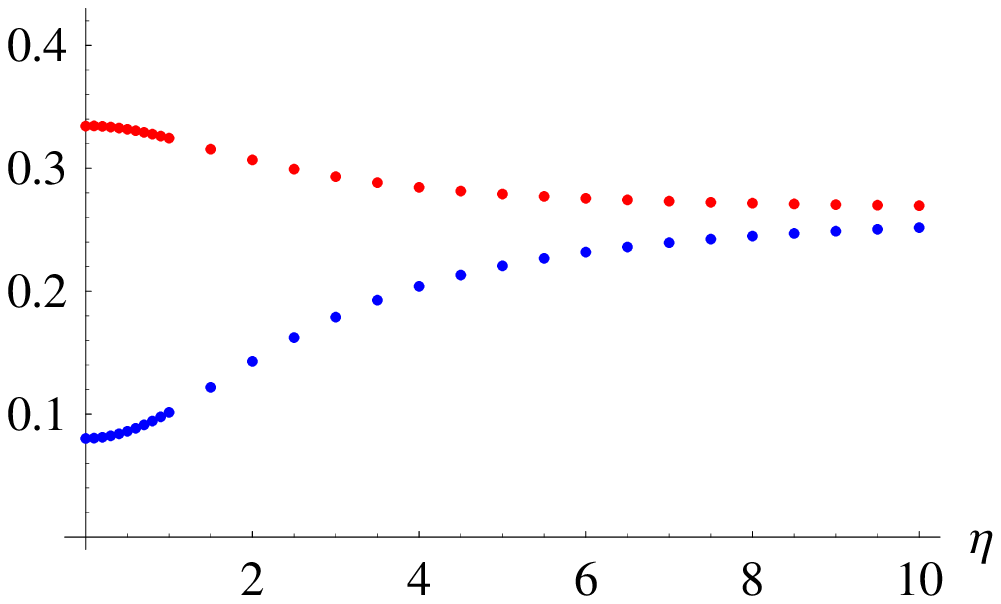}
&&
\includegraphics[width=5cm]{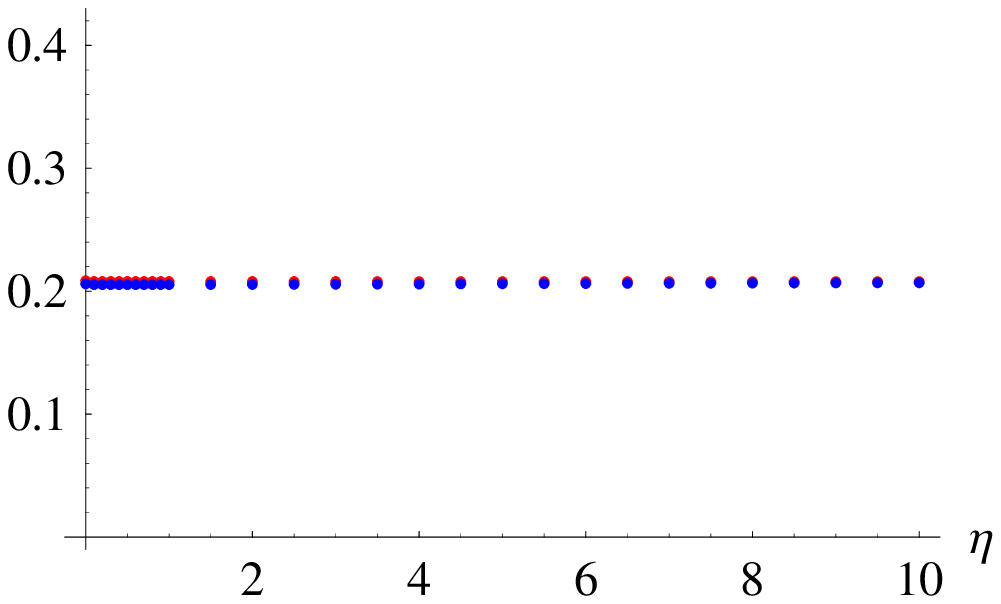}\\
$d=0$ && $d=1$ && $d=4$
\end{tabular}
\caption{{\small Effective potential vs. $\eta$. Red dots denote ${\cal V}_e$ while blue denote ${\cal
V}_g$.}}\label{fig:effv_e=g new}
\end{center}
\end{figure}

In Sec.~4 we have discussed the effective potential for two vortices for any separation and with any relative
orientation in a model with $\gamma=1$ and $\lambda = \lambda_g=\lambda_e$. The potential is shown in
Fig.~\ref{fig:effv_sep}. The effective potential comes from two pieces: the Abelian ${\cal V}_e$ and the non-Abelian
${\cal V}_g$ potential given in Eq.(\ref{potentials}). In Figs.~\ref{fig:effv_e=g} and \ref{fig:effv_e=g new}
we show ${\cal V}_e$ and ${\cal V}_g$ taking several  slices of
Fig.~\ref{fig:effv_sep}.

\begin{figure}[ht]
\begin{center}
\begin{tabular}{cc}
\includegraphics[width=8cm]{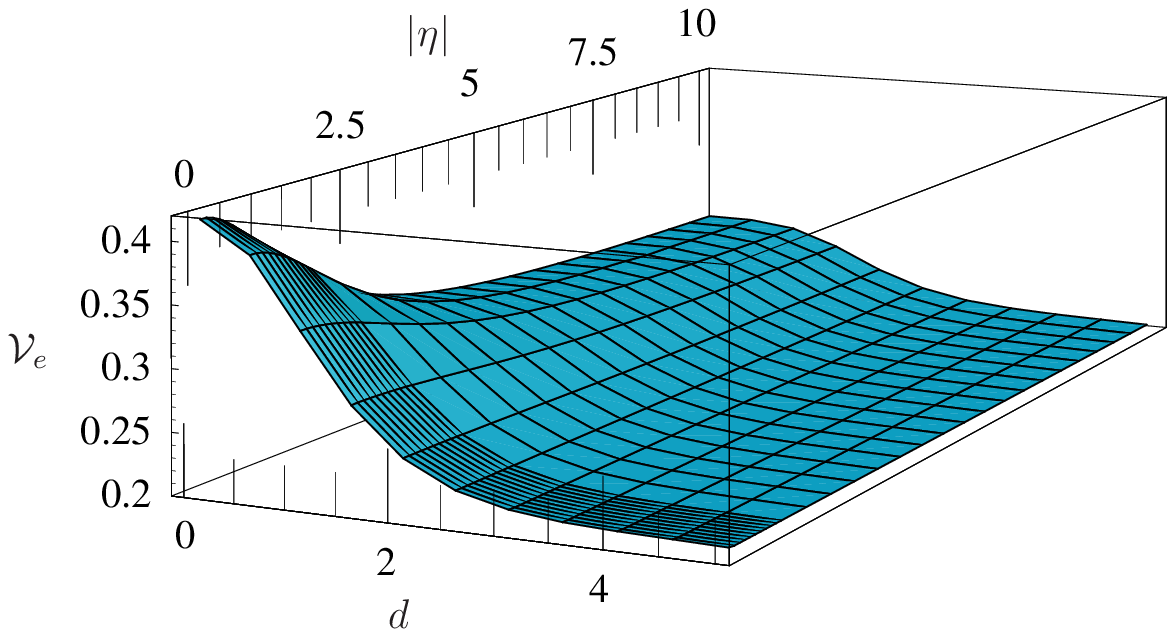} &
\includegraphics[width=8cm]{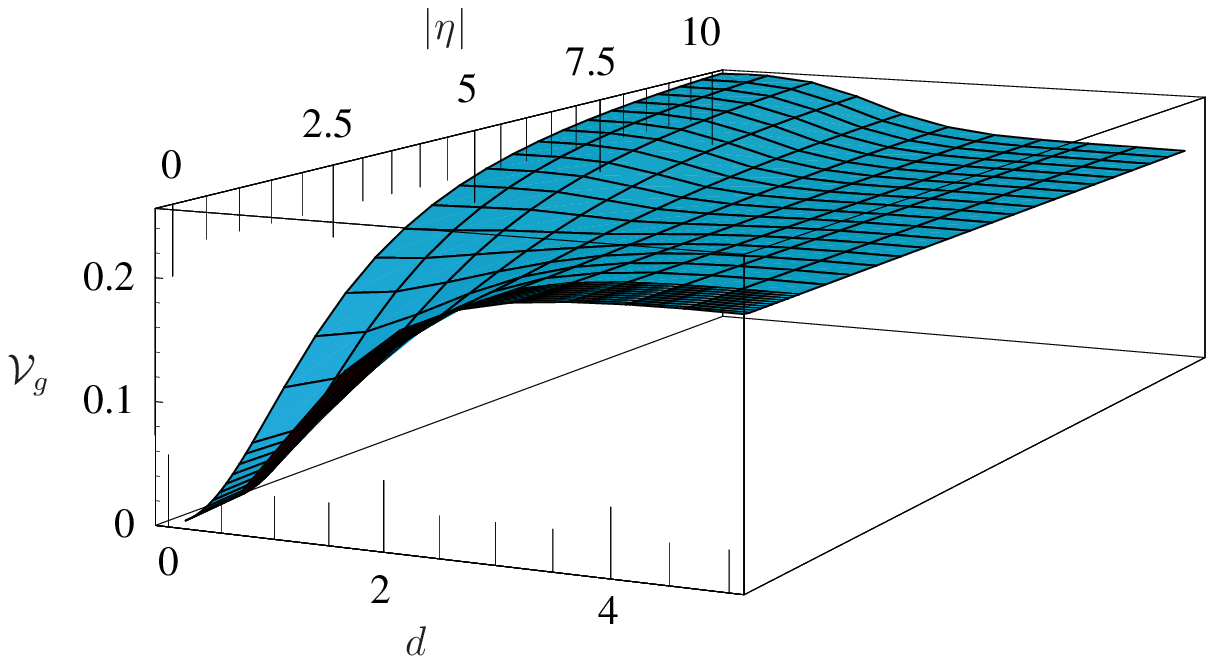}
\end{tabular}
\caption{{\small The Abelian potential ${\cal V}_e$ (left) and the non-Abelian potential ${\cal V}_g$ (right) for
$\gamma=1$.}} \label{fig:effv3d}
\end{center}
\end{figure}
Let us consider the case with $\lambda_e^2-1 > 0$ and $\lambda_g^2-1>0$. In this case the effective potential will have
the same qualitative behaviors like the reduced potentials in the Figs.~\ref{fig:effv_e=g}, \ref{fig:effv_e=g new}
and \ref{fig:effv3d}. The figures shows how ${\cal V}_e$ and ${\cal V}_g$ behaves very differently. In particular, the
Abelian potential is always repulsive, both in the real and internal space\footnote{Attraction in the internal space
here just means that orientations tend to become the same.} (see the red dots in
Figs.~\ref{fig:effv_e=g} and \ref{fig:effv_e=g new}). The non-Abelian potential is on the contrary sensitive on the
orientations.
In particular, Fig.~\ref{fig:effv_e=g} shows that it is repulsive for parallel vortices while it is
attractive for anti-parallel ones. Furthermore, the non-Abelian potential becomes almost flat (along the spatial
coordinate $d$) with the orientation $\eta \sim 4$, see the middle of Fig.~\ref{fig:effv_e=g}.
The blue dots in Fig.~\ref{fig:effv_e=g new}
reveal that the non-Abelian potential always gives attractive forces in the internal space. When the two scalar
couplings are equal, $\lambda_e^2=\lambda_g^2$, the left picture in Fig.~\ref{fig:effv_e=g} clearly shows how the two
potentials exactly cancel for anti-parallel vortices, recovering the result of the previous section.
\begin{table}[ht]
\begin{center}
\begin{tabular}{c||c|c|c}
 & $\lambda_g^2 > 1$ & $\lambda_g^2=1$ & $\lambda_g^2<1$\\
\hline \hline $\lambda_e^2>1$ &
\begin{minipage}{1.8cm}
\vspace{.1cm}
N=$(-,+)$\\
\vspace{-1cm}\\\noindent A=$+$ \vspace{.1cm}
\end{minipage}
&
\begin{minipage}{1.1cm}
\vspace{.1cm}
N=$0$\\
\vspace{-1cm}\\\noindent A=$+$ \vspace{.1cm}
\end{minipage}
&
\begin{minipage}{1.8cm}
\vspace{.1cm}
N=$(+,-)$\\
\vspace{-1cm}\\\noindent A=$+$ \vspace{.1cm}
\end{minipage}
\\
\hline $\lambda_e^2=1$ &
\begin{minipage}{1.8cm}
\vspace{.1cm}
N=$(-,+)$\\
\vspace{-1cm}\\\noindent A=$0$ \vspace{.1cm}
\end{minipage}
&
\begin{minipage}{1.1cm}
\vspace{.1cm}
N=$0$\\
\vspace{-1cm}\\\noindent A=$0$ \vspace{.1cm}
\end{minipage}
&
\begin{minipage}{1.8cm}
\vspace{.1cm}
N=$(+,-)$\\
\vspace{-1cm}\\\noindent A=$0$ \vspace{.1cm}
\end{minipage}
\\
\hline $\lambda_e^2<1$ &
\begin{minipage}{1.8cm}
\vspace{.1cm}
N=$(-,+)$\\
\vspace{-1cm}\\\noindent A=$-$ \vspace{.1cm}
\end{minipage}
&
\begin{minipage}{1.1cm}
\vspace{.1cm}
N=$0$\\
\vspace{-1cm}\\\noindent A=$-$ \vspace{.1cm}
\end{minipage}
&
\begin{minipage}{1.8cm}
\vspace{.1cm}
N=$(+,-)$\\
\vspace{-1cm}\\\noindent A=$-$ \vspace{.1cm}
\end{minipage}
\end{tabular}
\caption{{\small The forces between two vortices in the $\gamma=1$ case. N and A stand for non-Abelian force and
Abelian force, respectively. 0 means no force, $+$ means repulsive and $-$ means attractive. N=$(-,+)$ means that there
is an attractive force for anti-parallel vortices and a repulsive force for parallel vortices. }} \label{tab:gamma=1}
\end{center}
\end{table}

Of course, the true effective potential depends on $\lambda_e$ and $\lambda_g$ through the combination in
Eq.~(\ref{eq:gene_effpot}). This indicates the interaction between non-Abelian vortices is quite rich in comparison
with that of the ANO vortices.
Let us make a further example. The rightmost panel in  Fig.~\ref{fig:effv_e=g} shows that
${\cal V}_e$ and ${\cal V}_g$ for parallel vortices ($\eta=\infty$) are identical\footnote{This statement can be proved
analytically.}. Thus if we take $\lambda_e^2+\lambda_g^2=2$, the  interactions between parallel vortices vanishes while
between anti-parallel vortices they do not canceled out. This is just opposite to what we found in the case
$\lambda_e^2=\lambda_g^2$. We summarize the possible behaviors of the Abelian and non-Abelian forces, when the gauge
couplings are equal $e=g$ ($\gamma=1$), in Table~\ref{tab:gamma=1}.

\subsection{Different gauge coupling  $\gamma \neq 1$}

\begin{figure}[ht]
\begin{center}
\begin{tabular}{ccccc}
\includegraphics[width=5cm]{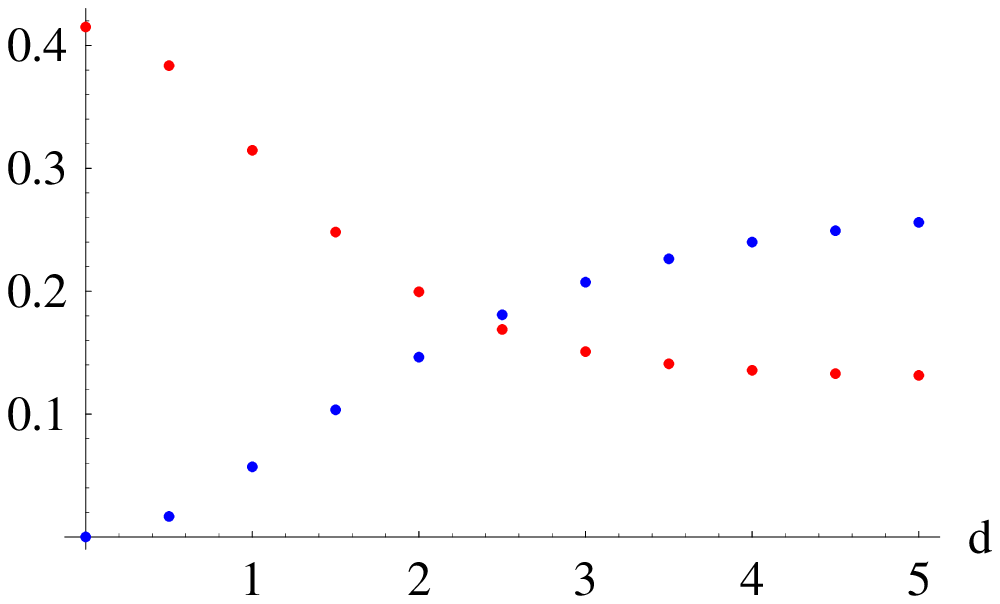}
&&
\includegraphics[width=5cm]{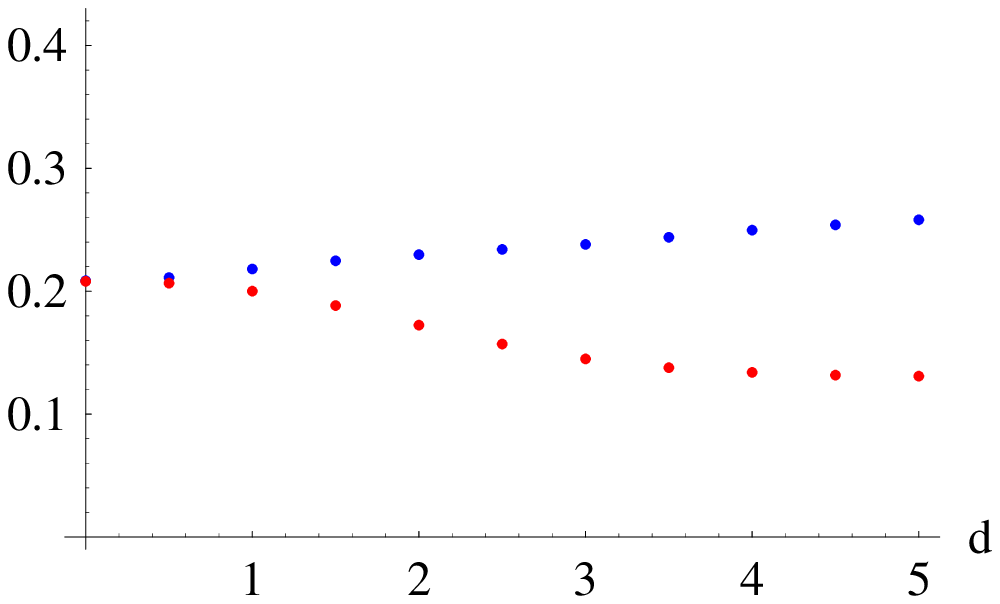}
&&
\includegraphics[width=5cm]{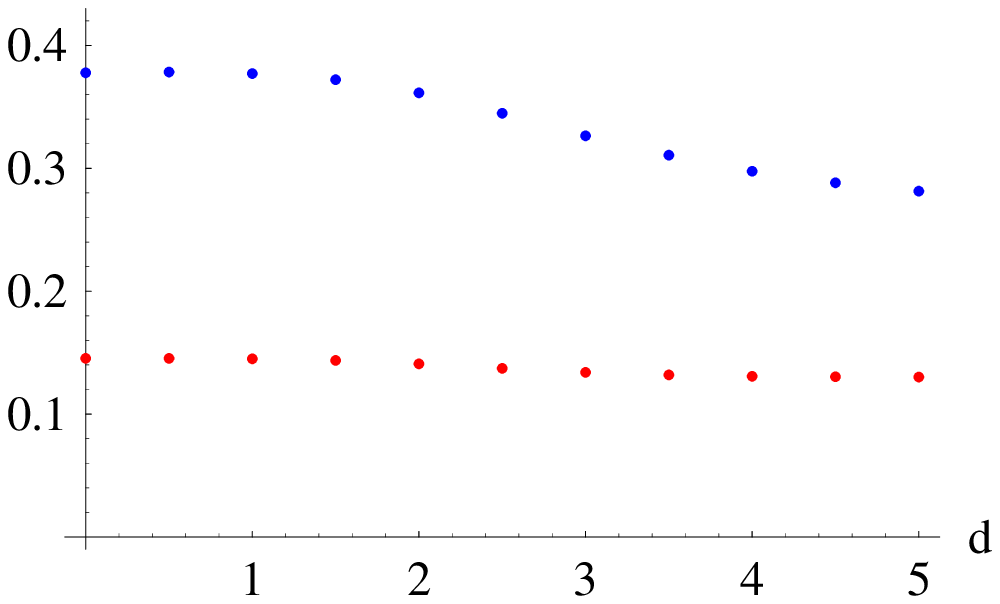}\\
anti-parallel ($\eta=0$) && intermediate ($\eta=4$) && parallel ($\eta=\infty$)
\end{tabular}
\caption{{\small Effective potential with $\gamma=1/2$ vs. separation. (red, blue) = (${\cal V}_e$, ${\cal V}_g$). }}
\label{fig:effv_e=2g}
\ \\
\begin{tabular}{ccccc}
\includegraphics[width=5cm]{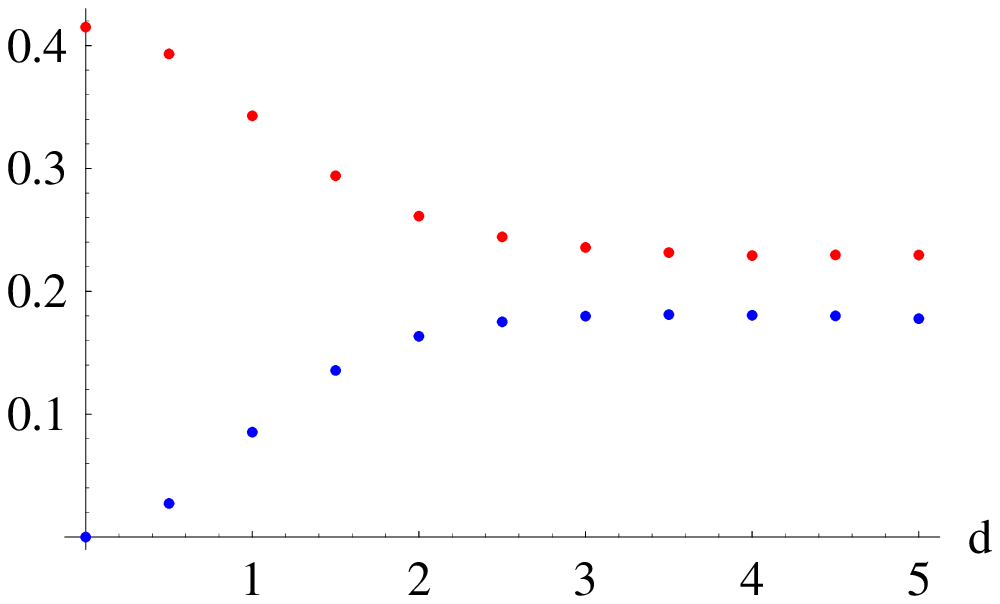}
&&
\includegraphics[width=5cm]{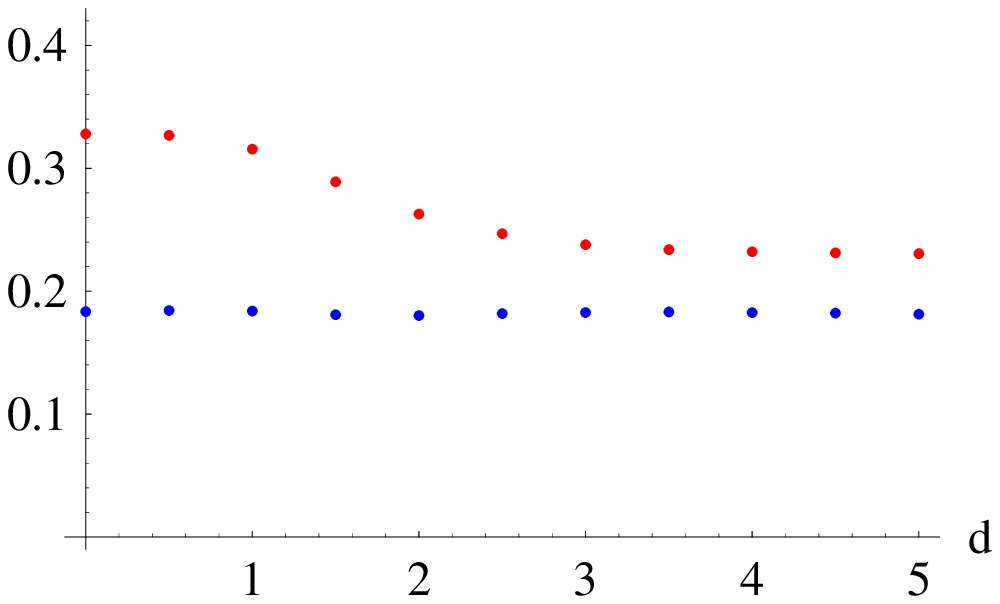}
&&
\includegraphics[width=5cm]{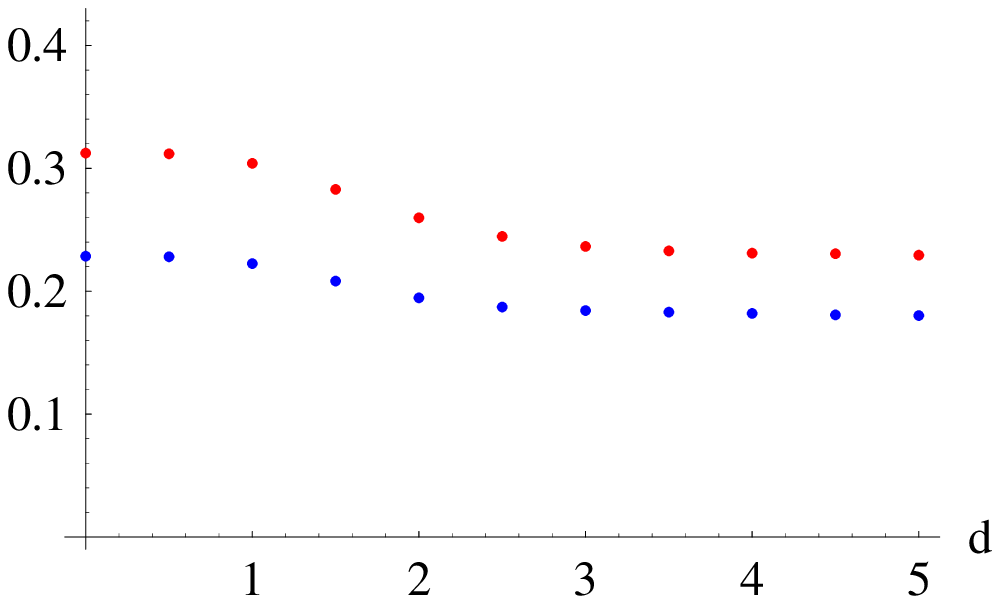}\\
anti-parallel ($\eta=0$) && intermediate ($\eta=4$) && parallel ($\eta=\infty$)
\end{tabular}
\caption{{\small Effective potential with $\gamma=1.3$ vs. separation. (red, blue) = (${\cal V}_e$, ${\cal V}_g$). }}
\label{fig:effv_g13}
\end{center}
\end{figure}

We now consider interactions between non-Abelian vortices with different gauge coupling $e\neq g$ ($\gamma \neq 1$).
Several numerical solutions are given in the left panel of Fig.~\ref{fig:numerics}.
In Figs.~\ref{fig:effv_e=2g} and \ref{fig:effv_g13} we show two numerical examples for the reduced effective potentials
${\cal V}_e$, ${\cal V}_g$
given in Eq.~(\ref{potentials}).

The plots show that the qualitative features of ${\cal V}_e$ and ${\cal V}_g$ are basically the same as
what is discussed
in the equal gauge coupling case $(\gamma=1)$.  Therefore, the qualitative
classification of the forces given in Table~\ref{tab:gamma=1} is still valid for $\gamma\neq1$.
We observe that the Abelian potential tends, at large distances, to a value
smaller than the non-Abelian one for $\gamma<1$, while opposite happens for $\gamma>1$.
The only things that have
non dependence on $\gamma$ are the values of the potentials at $(1,1)$-vortices with $(d,\eta)=(0,0)$.
Regardless of the gauge couplings ${\cal V}_g(\eta=0,d=0)=0$ while ${\cal V}_e(\eta=0,d=0) \fallingdotseq 0.41$.
This is because the
corresponding solution is proportional to the unit matrix, so that there are no contributions from the non-Abelian
part.

The effective potential is obtained from the linear combination in  Eq.~(\ref{eq:gene_effpot}) and
also depends on three parameters $\gamma$, $\lambda_e$ and $\lambda_g$. With this big freedom we can obtain a lot
of interesting interactions. For example, we can have potentials which develop a global minimum at some finite non zero
distance. In such cases two vortices may be bounded at that distance.
\begin{figure}[ht]
\begin{center}
\includegraphics[width=8cm]{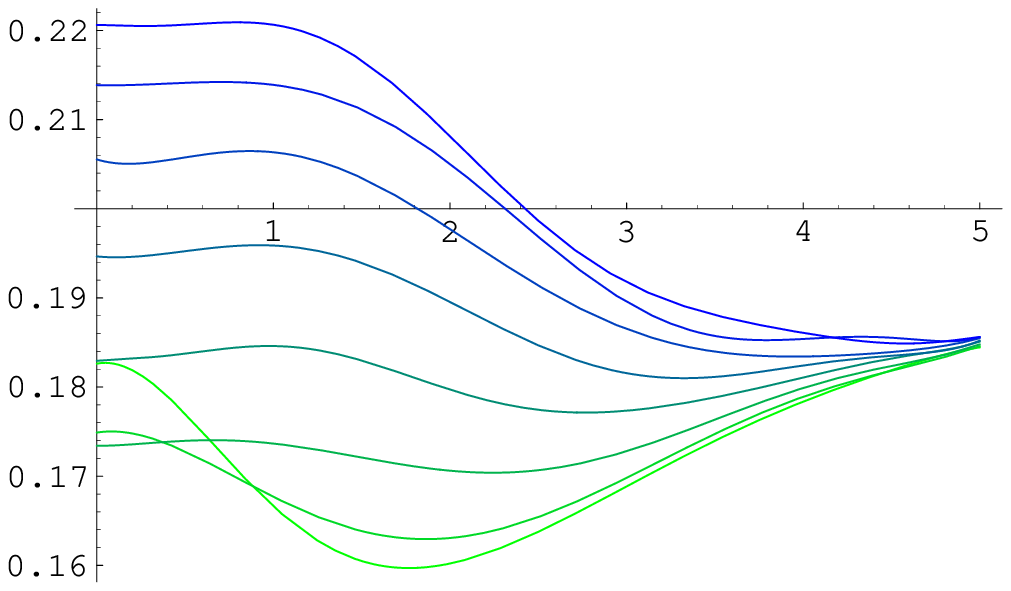}
\caption{{\small $\gamma=1/2$, $\lambda_e=1.2$, $\lambda_g=1.06$: From $\eta=0$ (green) to $\eta=7$ (blue) with
$d=0\sim5$ for each $\eta$.}}\label{bound}
\end{center}
\end{figure}
We can show a concrete potential in Fig.~\ref{bound}. The figure shows the presence of a minimum around $d\sim
2.$\footnote{These kind of potentials are really generic. In fact, the presence of this minima does not require strong
constraints on the couplings.} This kind of behavior have not been found for the ANO type I/II vortices and the
possibility of bounded vortices really results from the non-Abelian symmetry\footnote{Similar behaviors in the static
intervortex potential were also observed for $\mathbb{Z}_3$ vortices in \cite{Heo:1998ms}.}.

\section{Interaction at large vortex separation}

\subsection{Vortices in fine-tuned models $e=g$ and $\lambda_e=\lambda_g$ \label{sec:ana_int_fine}}

In this subsection we will obtain an analytic formula for the asymptotic forces between vortices
at large separation. We follow the technique developed in Refs. \cite{Speight:1996px,my}.
First of all, we need to find asymptotic behaviors of the scalar and the gauge fields.
We again consider the $(1,0)$-vortex
\beq
H_0(z)^{(1,0)} =
\left(
\begin{array}{cc}
z & 0\\
0 & 1
\end{array}
\right),\qquad
\vec\phi_1^{(1,0)} =
\left(
\begin{array}{cc}
1\\
0
\end{array}
\right).
\eeq
When we are sufficiently far from the core of the vortex, $Y(r)$ and $f(r)$
in Eqs.~(\ref{eq:reduced1}) and (\ref{eq:reduced2}) can be written as
\beq
Y = 2 \log r + \delta Y,\qquad
f = 1 + \delta f
\eeq
where $\delta Y$ and $\delta f$ are small quantities.
Plugging these into Eqs.~(\ref{eq:reduced1}) and (\ref{eq:reduced2}) and taking only
linear terms in $\delta Y$ and $\delta f$, we obtain the following linearized equations
\beq
&&\delta Y'' + \frac{1}{r}\delta Y' - \lambda^2 \delta Y - 2
\left( \delta f'' + \frac{1}{r}\delta f' - \lambda^2 \delta f \right) = 0,\\
&&\delta Y''' + \frac{1}{r}\delta Y'' - \frac{1}{r^2} \delta Y' - \delta Y' = 0.
\eeq
Solutions to these equations are analytically obtained to be
\beq
\delta Y - 2 \delta f = -\frac{q}{\pi} K_0(\lambda r),\quad
\delta Y = \frac{m}{\pi} K_0(r) - C,
\label{eq:asym_ano}
\eeq
where $K_0(r)$ is the modified Bessel's function of zeroth order and $q,m$ and $C$ are integration constants. $C$ must
be 0 because $\delta Y \to 0$ as $r \to \infty$ while $q,m$ should be determined by the original equation of motion. In
the BPS case, $\delta f \equiv 0$ ($f\equiv 1$), thus $q=-m$. Eq.~(\ref{eq:asym_ano})
leads to the well known asymptotic behavior of the
ANO vortex
\beq
H_{[1,1]} &=& f e^{-\frac{1}{2}Y}z
\simeq \left( 1 + \delta f - \frac{1}{2} \delta Y \right) e^{i\theta}
= \left( 1 + \frac{q}{2\pi}K_0(\lambda r) \right) e^{i\theta},
\label{eq:asymH}\\
\bar W_{[1,1]} &=& - \frac{i}{2} \bar \p Y \simeq - \frac{i}{4} e^{i\theta} \frac{d}{dr} \left( 2 \log r + \delta
Y\right) = - \frac{i}{2} \left(\frac{1}{r} - \frac{m}{2\pi}K_1(r) \right) e^{i\theta}, \label{eq:asymW}
\eeq
where $K_1 \equiv - K'_0$ and
we have defined $H_{[1,1]}$ and $\bar W_{[1,1]}$ as $[1,1]$ elements of
$H$ and $\bar W$ in Eq.~(\ref{eq:decomposition}) with the $k=1$ ansatz (\ref{eq:mm_para}).

Next we treat the vortices as point particles in a linear field theory
coupled with a scalar source $\rho$ and a vector current $j_\mu$.
To linearize the Yang-Mills-Higgs Lagrangian, we choose a gauge such that
the Higgs fields is given by the following hermitian matrix
\beq
H =
\left(
\begin{array}{cc}
1 & 0\\
0 & 1
\end{array}
\right)
+ \frac{1}{ 2}
\left(
\begin{array}{cc}
h^0+h^3 & h^1-ih^2 \\
h^1+ih^2 & h^0-h^3
\end{array}
\right),\quad
W_\mu = \frac{1}{ 2}
\left(
\begin{array}{cc}
w_\mu^0+w_\mu^3 & w_\mu^1 - iw_\mu^2\\
w_\mu^1+iw_\mu^2 & w_\mu^0-w_\mu^3
\end{array}
\right).
\label{eq:fluc}
\eeq
Here all the fields $h^a,w_\mu^a$ are real.
Then the quadratic part of the Lagrangian (\ref{eq:reL}) is of the form
\beq
{\cal L}^{(2)}_{\rm free} = \sum_{a=0}^3\left[
- \frac{1}{4} f_{\mu\nu}^a f^{a\mu\nu} + \frac{1}{2} w^a_\mu w^{a\mu}
+ \frac{1}{2}\p_\mu h^a \p^\mu h^a - \frac{\lambda^2}{2}(h^a)^2
\right]
\eeq
where we have defined the Abelian field strength $f_{\mu\nu}^a \equiv \p_\mu w^a_\nu - \p_\nu w^a_\mu$. We also take into
account the external source terms to realize the point vortex
\beq
{\cal L}_{\rm source} = \sum_{a=0}^3 \left[ \rho^a h^a - j_\mu^a w^{a\mu} \right]. \label{sources}
\eeq
The scalar and the vector sources should be determined so that the asymptotic behavior of the fields in
Eqs.~(\ref{eq:asymH}) and (\ref{eq:asymW}) are replicated. Equations of motions are of the form
\beq
\left(\square + \lambda^2\right)h^a = \rho^a,\qquad
\left(\square + 1\right)w_\mu^a = j_\mu^a . \label{sources2}
\eeq

In order to replicate the $(1,0)$-vortex corresponding to the $k=1$ ansatz (\ref{eq:mm_para}),
we just need to mimic the result of Refs.~\cite{Speight:1996px,my}
because the single non-Abelian vortex is a
mere embedding of the ANO vortex as mentioned earlier. In fact, only
$(h^0,w_\mu^0,\rho^0,j_\mu^0)=(h^3,w_\mu^3,\rho^3,j_\mu^3)$
are relevant and all the others are zero:
\beq
h^0=h^3 = \frac{q}{2\pi} K_0(\lambda r),&&\
\nonumber \rho^0=\rho^3 = q \delta(r),\\
{\bf w}^0={\bf w}^3 = - \frac{m}{2\pi} \hat{\bf k} \times \nabla K_0(r),&&\ {\bf j}^0 = {\bf j}^3= - m \hat{\bf k}
\times\nabla \delta(r)\label{tobedoubled}
\eeq
where $\hat{\bf k}$ is a spatial fictitious unit vector along the vortex world-volume. The vortex configuration with
general orientation is also treated easily, since the origin of the orientation is the Nambu-Goldstone mode associated
with the broken $SU(2)$ color-flavor symmetry
\beq
H_0 \to H_0(z)^{(1,0)} U_{\rm F},\quad \vec\phi_2 = U_{\rm F}^\dagger \vec\phi_1^{(1,0)}
= \left(
\begin{array}{c}
\alpha^*\\
\beta^*
\end{array}
\right),\quad
U_{\rm F} \equiv \left(
\begin{array}{cc}
\alpha & \beta \\
-\beta^* & \alpha^*
\end{array}
\right),
\label{eq:2nd_ori}
\eeq
where  $|\alpha|^2 + |\beta|^2 = 1$.
The fields $H$ and $W_\mu$ receive the following transformations, keeping
the hermitian form of (\ref{eq:fluc}).
\beq
\left(
\begin{array}{cc}
X & 0 \\
0 & 0
\end{array}
\right)
\to
U^\dagger_{\rm F} \left(
\begin{array}{cc}
X & 0 \\
0 & 0
\end{array}
\right)
U_{\rm F} =
\left(
\begin{array}{cc}
|\alpha|^2 & \alpha^*\beta \\
\alpha\beta^* & |\beta|^2
\end{array}
\right) X.
\label{eq:x}
\eeq
The scalar interaction between a vortex at ${\bf x} = {\bf x}_1$ with the orientation $\vec\phi_1$ and another vortex at
${\bf x} = {\bf x}_2$ with the orientation $\vec\phi_2$ can be obtained by
\beq
L_h &=&
\int dx^2\ \Tr\left[
\left(
\begin{array}{cc}
 h^0({\bf x}-{\bf x}_1) & 0 \\
0 & 0
\end{array}
\right)
\left(
\begin{array}{cc}
|\alpha|^2 & \alpha^*\beta \\
\alpha\beta^* & |\beta|^2
\end{array}
\right) \rho^0({\bf x} - {\bf x}_2)
\right] \nonumber\\
&=& |\alpha|^2 \frac{q^2}{2\pi} K_0(\lambda|{\bf x}_1 - {\bf x}_2|).
\label{eq:L_h}
\eeq
The gauge interaction is also obtained by similar way
\beq
L_w &=& -
\int dx^2\ \Tr\left[
\left(
\begin{array}{cc}
 {\bf w}^0({\bf x}-{\bf x}_1) & 0 \\
0 & 0
\end{array}
\right)\cdot
\left(
\begin{array}{cc}
|\alpha|^2 & \alpha^*\beta \\
\alpha\beta^* & |\beta|^2
\end{array}
\right) {\bf j}^{0}({\bf x} - {\bf x}_2)
\right] \nonumber\\
&=& - |\alpha|^2 \frac{m^2}{2\pi} K_0(|{\bf x}_1 - {\bf x}_2|).
\label{eq:L_w}
\eeq
Then total potential is $V_{\rm int} = - L_h - L_w$
\beq
V_{\rm int} = - \frac{\left|{\vec \phi}^\dagger_1{\vec \phi}_2\right|^2}{\left|{\vec \phi}_1\right|^2\left|{\vec
\phi}_2\right|^2} \left( \frac{q^2}{2\pi} K_0(\lambda|{\bf x}_1 - {\bf x}_2|) -\frac{m^2}{2\pi} K_0(|{\bf x}_1 - {\bf
x}_2|) \right), \label{eq:eff_e=g}
\eeq
where $|\alpha|^2=\frac{\left|{\vec \phi}^\dagger_1{\vec \phi}_2\right|^2}{\left|{\vec
\phi}_1\right|^2\left|{\vec \phi}_2\right|^2}$ is invariant under the global color-flavor rotation.
When two vortices have parallel orientations, this potential becomes that of two ANO
vortices \cite{Speight:1996px}. On the other hand, the potential vanishes when their orientations are
anti-parallel. This agrees with the numerical result found in the previous sections. In the BPS limit
$\lambda=1$ ($q=m$), the interaction becomes precisely zero.

Since $K_0(\lambda r)
\sim \sqrt{\pi/2\lambda r} e^{-\lambda r}$, the potential asymptotically reduces to
\beq
V_{\rm int} \simeq
\left\{
\begin{array}{lccl}
- \frac{\left|{\vec \phi}^\dagger_1{\vec \phi}_2\right|^2}{\left|{\vec \phi}_1\right|^2\left|{\vec \phi}_2\right|^2}
\dfrac{q^2}{2\pi} \sqrt{\dfrac{\pi}{2\lambda r}} e^{- \lambda r} &
 {\rm for} & \lambda < 1 & {\rm Type} \, {\rm I}\\
\frac{\left|{\vec \phi}^\dagger_1{\vec \phi}_2\right|^2}{\left|{\vec \phi}_1\right|^2\left|{\vec \phi}_2\right|^2}
 \dfrac{m^2}{2\pi} \sqrt{\dfrac{\pi}{2r}} e^{-r} & {\rm for} & \lambda > 1 & {\rm Type} \, {\rm II}
\end{array}
\right. \label{classif1}
\eeq
where $r \equiv |{\bf x}_1 - {\bf x}_2| \gg 1$. If we fix the relative orientation being some finite value, the force
$F_r = - \partial_r V_{\rm int}$ between two vortices is attractive for $\lambda < 1$ and repulsive for $\lambda > 1$
similar to the force between ANO vortices. The force vanishes when the relative orientation becomes anti-parallel.
If we fix the distance by hand, the orientations tend to be anti-parallel for the type II while the parallel
configuration are preferred for the type I case.

\subsection{Vortices with general couplings}

It is quite straightforward to generalize the results
of the previous section to the case of generic couplings.
We can of course use the same gauge as in Eq.~(\ref{eq:fluc}).
The quadratic Lagrangian  (\ref{flum}) is of the form
\beq
{\cal L}^{(2)}_{\rm free} &=& \sum_{a=1}^3\left[
- \frac{1}{4 \gamma^2} f_{\mu\nu}^a f^{a\mu\nu} + \frac{1}{2} w^a_\mu w^{a\mu}
+ \frac{1}{2}\p_\mu h^a \p^\mu h^a - \frac{\lambda_g^2 \gamma^2}{2}(h^a)^2
\right]\nonumber\\
&+& \left[
- \frac{1}{4} f_{\mu\nu}^0 f^{0\mu\nu} + \frac{1}{2} w^0_\mu w^{0\mu}
+ \frac{1}{2}\p_\mu h^0 \p^\mu h^0 - \frac{\lambda_e^2}{2}(h^0)^2
\right].
\eeq
The external sources can be still reproduced by source terms as in Eqs.~(\ref{sources}), (\ref{sources2}). The
linearized equations following from the above Lagrangian are of the form
\beq
\left(\frac{1}{\gamma^2}\square + 1\right) w_\mu^a = j_\mu^a,\quad
\left(\square + \lambda_g^2\gamma^2\right) h^a = \rho^a,\quad
\left(\square + 1\right) w_\mu^0 = j_\mu^0,\quad
\left(\square + \lambda_e^2\right) h^0 = \rho^0.
\eeq

For the $(1,0)$-vortex
the only non-zero profile functions are $(h^0,w_\mu^0,\rho^0,j_\mu^0)$ and $(h^3,w_\mu^3,\rho^3,j_\mu^3)$. But these
profiles are no longer equal and we need to deal with them independently. The only difference from the
similar equations (\ref{sources2}) is  for the masses of the particles. The masses are directly related to asymptotic
tails of vector and scalar fields. We can easily find the solutions by doubling Eqs.~(\ref{tobedoubled})
\beq
h^{0} = \frac{q^{0} }{2\pi} K_0(\lambda_e r), \quad
{\bf w}^{0}  = - \frac{m^{0} }{2\pi} \hat{\bf k} \times \nabla K_0(r),\quad
\rho^{0}  = q^{0}  \, \delta(r),\quad
{\bf j}^{0}  = - m^{0}  \, \hat{\bf k} \times\nabla \delta(r),\\
h^{3} = \frac{q^{3} }{2\pi} K_0(\lambda_g\gamma r), \quad
{\bf w}^{3}  = - \frac{m^{3} }{2\pi} \hat{\bf k} \times \nabla K_0(\gamma r),\quad
\rho^{3}  = q^{3}  \, \delta(r),\quad
{\bf j}^{3}  = - m^{3}  \, \hat{\bf k} \times\nabla \delta(r).
\eeq
The vortex with the orientation $\vec\phi_2$ in Eq.~(\ref{eq:2nd_ori}) can be obtained
by performing an $SU(2)_{{\rm C}+{\rm F}}$ rotation like Eq.~(\ref{eq:x})
\beq
&&\left(
\begin{array}{cc}
\frac{X^0+X^3}{2} & 0 \\
0 & \frac{X^0-X^3}{2}
\end{array}
\right)\nonumber\\
&&\rightarrow
\left(
\begin{array}{cc}
\frac{X^0}{2} + \left(|\alpha|^2 - |\beta|^2\right)\frac{X^3}{2} &
\left(\alpha^*\beta-\beta\alpha^*\right)\frac{X^0}{2} + \left(\alpha^*\beta+\beta\alpha^*\right)\frac{X^3}{2} \\
\left(\alpha\beta^*-\beta^*\alpha\right)\frac{X^0}{2} + \left(\alpha^*\beta+\beta\alpha^*\right)\frac{X^3}{2}
& \frac{X^0}{2} - \left(|\alpha|^2 - |\beta|^2\right)\frac{X^3}{2}
\end{array}
\right).
\eeq
Similar to Eqs.~(\ref{eq:L_h}) and (\ref{eq:L_w}), we find
the total potential $V_{\rm int}$
\beq
V_{\rm int} &=& \frac{1}{2}\left(
-\frac{(q^0)^2}{2\pi} K_0(\lambda_e |{\bf x}_1 - {\bf x}_2|)
+\frac{(m^0)^2}{2\pi} K_0(|{\bf x}_1 - {\bf x}_2|) \right)\nonumber\\
&+& \left(\frac{\left|{\vec \phi}^\dagger_1{\vec \phi}_2\right|^2}{\left|{\vec \phi}_1\right|^2\left|{\vec
\phi}_2\right|^2} - \frac{1}{2}\right) \left( -\frac{(q^3)^2}{2\pi} K_0(\lambda_g \gamma |{\bf x}_1 - {\bf x}_2|)
+\frac{(m^3)^2}{2\pi} K_0(\gamma |{\bf x}_1 - {\bf x}_2|) \right).
\eeq
When we tune the parameters to $\gamma=1$ and $\lambda_g=\lambda_e$
($q^0=q^3,\ m^0 = m^3$), this
effective potential is exactly identical to that of Eq.~(\ref{eq:eff_e=g}).
In the BPS limit $\lambda_e=\lambda_g=1$, the interaction becomes precisely zero
because $q^0=m^0$ and $q^3=m^3$.

At large distance, the interactions between vortices are dominated by the particles with the lowest mass $M_{\rm low}$.
There are four possible regimes
\beq
V_{\rm int} =
\left\{
\begin{array}{ccll}
-  \dfrac{(q^0)^2}{4\pi} \sqrt{\dfrac{\pi}{2 \lambda_e r}}
 e^{- \lambda_e r}  & {\rm for} & M_{\rm low}=M_{\rm s},& {\rm Type} \, {\rm I}\\
 - \left(\frac{\left|{\vec \phi}^\dagger_1{\vec \phi}_2\right|^2}{\left|{\vec \phi}_1\right|^2\left|{\vec
\phi}_2\right|^2} - \frac{1}{2}\right)
   \dfrac{(q^3)^2}{2\pi} \sqrt{\dfrac{\pi}{2\lambda_g \gamma r}}
 e^{-  \lambda_g \gamma r}  & {\rm for} & M_{\rm low}=M_{\rm ad},& {\rm Type} \, {\rm I}^* \\
 \dfrac{(m^0)^2}{4\pi} \sqrt{\dfrac{\pi}{2 r}}
 e^{- r} & {\rm for} & M_{\rm low}=M_{U(1)},& {\rm Type} \, {\rm II}\\
 \left(\frac{\left|{\vec \phi}^\dagger_1{\vec \phi}_2\right|^2}{\left|{\vec \phi}_1\right|^2\left|{\vec
\phi}_2\right|^2} - \frac{1}{2}\right)
 \dfrac{(m^3)^2}{2\pi} \sqrt{\dfrac{\pi}{2 \gamma r}}
  e^{- \gamma r}  & {\rm for} & M_{\rm low}=M_{SU(2)},&{\rm Type} \, {\rm II}^*
\end{array}
\right. \label{classif2}
\eeq
which exhaust all the possible kinds of asymptotic potentials of this system. This generalizes the type I/II
classification of Abelian superconductors. We have found two new categories, called type I$^*$ and type II$^*$, in
which the force can be attractive or repulsive depending on the relative orientation. The type I force is always
attractive and the type II force is repulsive regardless of the relative orientation. On the other hand, type I$^*$ and
type II$^*$ depend on the relative orientation. In the type I$^*$ case the forces between parallel vortices are
attractive while anti-parallel vortices repel each other. The type II$^*$ vortices feel opposite forces to the type
I$^*$\footnote{In Ref.~\cite{1stpaper} we have found a similar result in a supersymmetric theory. The relation between
the two notations is explained in the Appendix. The supersymmetric theory in Ref.~\cite{1stpaper} shows only type
I/I$^*$ behaviors.}. Note that we have used the same terms type I/II for the fine-tuned model in
sect.~\ref{sec:ana_int_fine}. In the perspective of this section, they should be called type I$+$I$^*$/II$+$II$^*$
because of the degeneracy of some masses. It is interesting to compare these results with the recently
studied asymptotic interactions between non-BPS non-Abelian global vortices \cite{global}. A very different feature of
global vortices is that the interactions are always repulsive. This is because they are mediated by Nambu-Goldstone
zero modes, whereas in our model these particles are all eaten by the gauge bosons thanks to the Higgs mechanism.

We find a nice matching of qualitative features
between the numerical results of the previous sections and the semi-analytical
results in this section. Let us look at Figs.~\ref{fig:effv_e=g}, \ref{fig:effv_e=2g} and \ref{fig:effv_g13}.
In all the cases, we found that the Abelian potentials are attractive regardless of the orientations
while the non-Abelian potentials are sensitive to those. These properties are well shown also in
the semi-analytical results in (\ref{classif2}). The type I/II interactions originated by the $U(1)$ part
are independent of the orientations
whereas the type I$^*$/II$^*$ which are coming from the $SU(N)$ part do depend on them.

The result in Eq.~(\ref{classif2}) is easily extended to the general case of $U(1) \times SU(N)$.
This can be done by just thinking of the orientation vectors $\vec \phi$ as taking values
in ${\bf C}P^{N-1}$.

\section{Conclusion and discussion}

In this paper we have studied static interactions between non-BPS vortices in $SU(N) \times U(1)$ gauge theories with
Higgs fields in the fundamental representation. We have discussed models with arbitrary gauge and scalar couplings. We
have numerically computed the effective potential for almost BPS configurations for arbitrary
separations and any internal orientations.
We have also obtained analytic expressions for the static forces between well separated non-BPS vortices.
This expression is valid also for models far from BPS limit.

For the fine-tuned model we found interaction pattern similar to that of the ANO vortices in the
Abelian-Higgs model. The numerical effective potential is given in Fig.~\ref{fig:effv_sep}, it depends on both the
relative distance and the orientations. The asymptotic potential between two vortices is given by Eq.~(\ref{classif1}).
In this model the mass of the $U(1)$ and of the $SU(N)$ vector bosons are same $M_{SU(N)}= M_{U(1)}$, and also all the
scalars have the same masses $M_{\rm s} = M_{\rm ad}$. We thus have only two mass scales, which corresponds to two
different asymptotic regimes. For $\lambda<1$ ($M_{\rm s} < M_{U(1)}$) there is universal attraction (type I) and for
$\lambda>1$ ($M_{\rm s}
> M_{U(1)}$) universal repulsion (type II). Both the numerical and the analytical result show that the interactions
between two anti-parallel vortices vanish; this configuration is unstable for type I vortices and stable
for type II. So in this last case the part of the moduli space which corresponds
to vortices with opposite ${\bf C}P^1$ orientations at arbitrary distance survives the
non-BPS perturbation.

In models with arbitrary couplings, on the other hand, the pattern of interactions becomes richer. In this case we
considered separately  the Abelian and non-Abelian contributions ${\cal V}_e$ and ${\cal V}_g$
to the effective potential.
The two show very different qualitative behavior. While the Abelian contribution is always attractive
(or repulsive) for a given choice of the parameters, the non-Abelian one can be attractive or repulsive
depending on the relative internal orientation of the two vortices. These properties combined with the fact that the
full effective potential is the linear combination given in Eq.~(\ref{eq:gene_effpot}) deduce that we can obtain
many qualitatively different type of interactions depending on the choice of the parameters of the theory. Such a variety
also appears in the possible asymptotic behavior of the interactions. In the theory there are four
different mass scales, the masses of the $SU(N)$ vector bosons $M_{SU(N)}$, the $U(1)$ vector boson $M_{U(1)}$, the
adjoint scalars $M_{\rm ad}$ and the singlet scalar $M_{\rm s}$ under the color-flavor symmetry.
This leads to four different asymptotic regimes, classified in
Eq.~(\ref{classif2}). We found, in addition to type I and type II, new types of interaction mediated by the non-Abelian
particles, which we call type I$^*$ and type II$^*$. The type I (type II) force is attractive (repulsive),
and occurs when the singlet scalar ($U(1)$ vector boson)
has the smallest mass. These forces do not depend on the relative orientation. In the type I$^*$ case there is an
attractive force for parallel orientations and repulsive for anti-parallel ones where the asymptotic force is mediated
by the lightest non-Abelian scalar fields. On the other hand for type II$^*$ there is repulsion for parallel orientation and
attraction for anti-parallel ones and the force at large distance is mediated by the lightest non-Abelian vector field.

The dynamics of the interactions of the non-BPS non-Abelian vortices is quite rich.
Let us give comments on some possible further directions:
\begin{itemize}
\item {\it  Reconnection rate of the cosmic string:}
The slow moving non-Abelian BPS vortex strings, as cosmic strings,
 were shown to always reconnect
with probability one \cite{Hashimoto:2005hi,Eto:2006db}. This
is important in order to distinguish solitonic cosmic strings
from fundamental cosmic strings which generically have very small reconnection probability.
Consider the two dimensional space spanned by $\eta$ and $d$; the coordinate $d$ corresponds to
the vortices relative distance and $\eta$ to the internal colour-flavour orientation.
The $(2,0)$-vortices are at $\eta \rightarrow \infty$
and the $(1,1)$-vortices at $\eta=0$.
Only the orbits which pass through the point $d=0,\eta=0$
correspond to scatterings where no reconnection occurs;
these orbits represent scatterings with very finely tuned initial conditions.
The fine-tuned scattering process cannot contribute
to the reconnection probability; for this reason the reconnection rate
of the non-Abelian BPS vortex is one~\cite{Hashimoto:2005hi,Eto:2006db}.

When we consider strong non-BPS corrections, the moduli space approximation is no longer valid and this
conclusion could drastically change. For some values of the couplings
there will appear regimes
in which the coincident $(1,1)$-vortices are favored to the coincident $(2,0)$-vortices. In that case we may expect that
the reconnection probability becomes smaller than one. As a very simple example, we can consider the fine tuned
model for $\lambda>1$. In this case the energetically favored configuration is the one with two vortices with opposite
${\bf C}P^1$ orientation, which never reconnect. In this case we can expect a reduction of the reconnection rate. On
the contrary for $\lambda<1$, when the $(2,0)$ state is energetically favored, we expect that the reconnection rate
should still be one. It would be interesting to make a detailed numerical investigation of the scattering process of non-BPS
non-Abelian vortices, in order to clarify how the non-BPS corrections could modify the reconnection rate.

\item {\it  Non-Abelian vortices and Abrikosov lattice:}
In the usual type II superconductor (the Abelian-Higgs model), if a large number of vortices penetrate a region of
given area  $A$, they will form a hexagonal lattice (Abrikosov lattice) rather than forming a square one
\cite{Kleiner1964}. This is verified experimentally. If we consider the same for  non-Abelian  vortices, we expect that
the property of the lattice can be quite different, say, lattice spacing and/or form can change. An interesting
possibility is the appearance of phase transitions due to the change of the lattice structure when the density of
vortices change. This eventuality is suggested by the possible presence of interactions which change with distances. For
example in the case of Fig.~\ref{bound} there are repulsive forces at short distances while at large distances they are
attractive.

\item {\it  Quantum aspects:} In this paper we focused on the classical aspects of the interactions between
non-Abelian vortices. In the theoretical set-up that we have discussed, the quantum aspects of the infrared physics of
a single vortex are described by an effective bosonic ${\bf C}P^{N-1}$ sigma model
(the theoretical setting is in this sense similar  to the one discussed Ref.~\cite{nsgsy}). Let us
consider two vortices at large distance; in the type I$^*$ and in the type II$^*$ regimes the quantum physics will be
described by two  ${\bf C}P^{N-1}$ sigma models with an interaction potential given by Eq.~(\ref{classif2}). It
would be interesting to study the effect of this term in the sigma model physics. Another interesting problem is the
numerical determination of the effective theory for vortices  at generic separations. To do this one has to determine
the Manton metric on the full moduli space.

\end{itemize}

\section*{Acknowledgments}

We are grateful to Alexander Gorsky, Sven Bjarke Gudnason, Kenichi Konishi, Keisuke Ohashi, Giacomo Marmorini and
Muneto Nitta for useful discussions and comments. W.V.~wants to thanks Marco Tarallo for comments and discussions
about the hidden subtleties of the numerical calculations. The work of M.E.~is supported by the Research Fellowships of
the Japan Society for the Promotion of Science for Research Abroad.

\appendix

\section{Note on the relation between two formalisms}

The aim of this small section is to relate the moduli matrix formalism that we used in this paper to the direct ansatz
approach we choose in a previous work \cite{1stpaper}. To this end we must find the relation between the angle $\alpha$
used in \cite{1stpaper} and the moduli matrix parameters $\eta=1/a'$. The resulting relation between the two variables,
in the case of coincident vortices, is rather non trivial.
\begin{description}
  \item[Coincident vortices.] In the moduli matrix formalism, two coincident vortices are described by the following
  moduli matrix
  \beq
  H_0^{(1,1)} = \left(
  \begin{array}{cc}
  z & -\eta\\
  0 & z
  \end{array}
\right),
  \eeq
while in \cite{1stpaper} we used explicit ansatzs for the fields. In particular, for the squarks field we
have used
\beq
 H_{\rm ans} =  \left(\begin{array}{cc}
-\cos \frac{\alpha}{2}  e^{ 2 i \varphi}  \kappa_1(r)
& \sin \frac{\alpha}{2}  e^{  i \varphi} \kappa_2(r)  \\[3mm]
- \sin \frac{\alpha}{2}  e^{  i \varphi}  \kappa_3(r)
& -\cos \frac{\alpha}{2} \kappa_4(r)  \\
\end{array}\right). \label{ans1}
\eeq
This form for the squark fields lead us, in \cite{Auzzi:2005gr,Eto:2006cx}, to
conjecture that  this solution is associated with that given
by the following moduli matrix:
\beq
H_{\rm ans} \quad \Leftrightarrow \quad H_{0,{\rm ans}}=\left(
  \begin{array}{cc}
  -\cos \frac{\alpha}{2} z^2 & \sin \frac{\alpha}{2} z\\
  -\sin \frac{\alpha}{2} z & -\cos \frac{\alpha}{2}
  \end{array}
\right).
\eeq
Using a $V$ equivalence we can put $ H_{0,ans}$ on the standard $(1,1)$ form
\beq
 H_0^{(1,1)}=V(z) H_{0,{\rm ans}}=\left(
  \begin{array}{cc}
  z & \cot(\alpha/2)\\
  0 & z
  \end{array}
\right).
\eeq
This simple argument gave us the relation: $\eta=\cot(\alpha/2)$.

Here we point out that this conjecture is wrong. In fact the functions
$(\kappa_1,\kappa_2,\kappa_3,\kappa_4)$ in Eq.~(\ref{ans1}) have an implicit dependence on $\alpha$, which come out
only after solving the differential equations for the vortices.
This means that it is difficult to find a simple analytical relation
between the parameters of the two formalisms.
Also, this relation is not the same for all the values of $\gamma=g/e$.
In Fig.~\ref{fig:numrel} we plotted the numerical relation between
$\alpha$ and $\eta$ for $\gamma=1$.
 The corresponding values for the two variables are found making an empirical match of some gauge
invariant functions which have a non-trivial dependence on the relative orientation.
One of these functions,
$\Tr F_{12}$, is showed in Fig.~\ref{fig:mflux}.
The fact that a perfect matching is possible between gauge invariant functions in both the formalisms
is a strong numerical check of the consistence of the two approaches.

\begin{figure}[ht]
\begin{center}
\includegraphics[width=7cm]{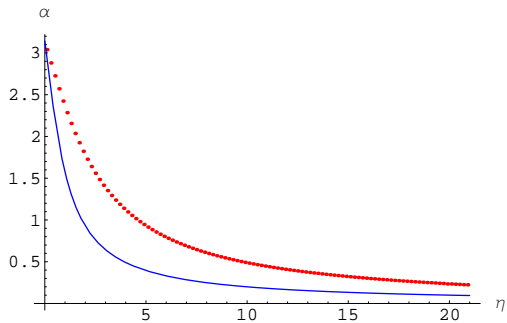}
\caption{{\small
The solid line is the naive relation $\eta=\cot(\alpha/2)$ and the dots are the true
numerical relations between $\eta$ and $\alpha$ for $\gamma=1$
}} \label{fig:numrel}
\end{center}
\end{figure}

  \item[Well separated vortices.]  Vortices at large separation have well-defined global orientations in the internal
  space, and thus a well-defined notion of relative orientation. This enables us
  to find an exact relation between the parameters of the two formalisms.
  In \cite{1stpaper} the global orientation of a non-Abelian vortex was defined by a vector which
  takes values in the internal space: $\vec n$. Using
  global color-flavor transformations we can always put the orientations
  ${\vec n}_1$ and ${\vec n}_2$ of two vortices on the following standard form
  \beq
  {\vec n}_1=(0,0,1), \quad {\vec n}_2=(-\sin\alpha,0,\cos\alpha).
  \eeq
  The vector ${\vec n}_2$ can be obtained from ${\vec n}_1$ acting with a global rotation
  \beq
  {\vec n}_2 \cdot \vec \tau \equiv U^{-1} \tau^3 U, \quad {\rm with} \quad U=\left(
  \begin{array}{cc}
  \cos \frac{\alpha}{2}  & -\sin \frac{\alpha}{2} \\
  \sin \frac{\alpha}{2}  & \cos \frac{\alpha}{2}
  \end{array}
\right).
  \eeq
We can repeat the same argument for the orientation vectors defined within the moduli matrix formalism (see
Eq.~(\ref{fundorient})). We can choose the following identification:
\beq
({\vec n}_1,{\vec n}_2)=((0,0,1),(-\sin\alpha,0,\cos\alpha)) \quad  \Leftrightarrow \quad ({\vec \phi}_1,{\vec
\phi}_2)=\left(\left(\begin{array}{c} 1 \\    0 \\\end{array} \right),\left(\begin{array}{c}1
\\b'\\\end{array}\right)\right).\eeq
If we act with the same global rotation on ${\vec \phi}_1$
\beq
{\vec \phi}_2=U^\dagger {\vec \phi}_1=\left(
                                           \begin{array}{c}
                                             \cos \frac{\alpha}{2}  \\
                                             -\sin \frac{\alpha}{2} \\
                                           \end{array}
                                         \right)\sim \left(
                                           \begin{array}{c}
                                            1   \\
                                             -\tan\frac{\alpha}{2} \\
                                           \end{array}
                                         \right),
\eeq
we find the relation $b'= -\tan\frac{\alpha}{2}$. With this identification we can easily check the identity:
\beq
\frac{\left|{\vec \phi}^\dagger_1{\vec \phi}_2\right|^2}{\left|{\vec \phi}_1\right|^2\left|{\vec
\phi}_2\right|^2}-\frac{1}{2}=\frac{{\vec n}_1\cdot{\vec n}_2}{2}=\frac{\cos\alpha}{2}.
\eeq
This shows the consistence of the expressions for the asymptotic forces obtained in this paper in
Eq. (\ref{classif2}) with those found in \cite{1stpaper}.

\end{description}


\end{document}